\documentclass[%
 reprint,
superscriptaddress,
nofootinbib,
 amsmath,amssymb,
 aps,
prd,
floatfix,
]{revtex4-2}
\usepackage{CJKutf8}

\usepackage{amsmath}
\usepackage{graphicx}%
\usepackage{dcolumn}%
\usepackage{bm}%
\usepackage[table,svgnames,dvipsnames]{xcolor}
\usepackage[mathlines]{lineno}
\usepackage[normalem]{ulem}
\usepackage{aas_macros}
\usepackage{graphicx}%
\usepackage{graphics}
\usepackage{dcolumn}%
\usepackage{bm}%
\usepackage{cases}%
\usepackage{booktabs}
\usepackage{comment}
\usepackage{multirow}
\usepackage{makecell}
\usepackage{siunitx}
\usepackage{tabularx}
\usepackage{xspace}
\usepackage{soul} %
\usepackage{fontawesome}

\usepackage{orcidlink} %
\usepackage{hanging} %

\def\be{\begin{equation}}
\def\ee{\end{equation}}
\def\ba#1\ea{\begin{align*}#1\end{align*}}

\newcommand{\lisw}{\mathrm{LISW}}
\newcommand{\Mpc}{\, \text{Mpc}}
\newcommand{\Gpc}{\, \text{Gpc}}
\usepackage{hyperref}
\usepackage[nameinlink,noabbrev]{cleveref}
\crefname{equation}{Eq.}{Eqs.}
\crefname{section}{Section}{Sections}
\crefname{figure}{Figure}{Figures}
\crefname{table}{Table}{Tables}
\crefname{appendix}{Appendix}{Appendices}
\Crefname{figure}{Figure}{Figures}
\Crefname{equation}{Equation}{Equations}
\Crefname{section}{Section}{Sections}
\Crefname{table}{Table}{Tables}
\definecolor{linkblue}{rgb}{0.0,0.25,0.6}
\hypersetup{colorlinks=true, linkcolor=linkblue, citecolor=linkblue, urlcolor=linkblue}

\begin{document}
\begin{CJK*}{UTF8}{gbsn}

\title{Acoustic priors from the latest cosmic microwave background data}

\author{Kushal Lodha\orcidlink{0009-0004-2558-5655}}
\email{klodha@bu.edu}
\affiliation{Korea Astronomy and Space Science Institute, 776, Daedeokdae-ro, Yuseong-gu, Daejeon 34055, Republic of Korea}
\affiliation{University of Science and Technology, 217 Gajeong-ro, Yuseong-gu, Daejeon 34113, Republic of Korea}
\affiliation{Department of Physics, Boston University, Boston, MA 02215, USA}

\author{Jun-Qian Jiang (江俊钱)\orcidlink{0000-0003-0957-3633}}
\email{jiang@kasi.re.kr}
\affiliation{Korea Astronomy and Space Science Institute, 776, Daedeokdae-ro, Yuseong-gu, Daejeon 34055, Republic of Korea}

\author{Purba Mukherjee\orcidlink{0000-0002-2701-5654}}
\email{purba@kasi.re.kr}
\affiliation{Korea Astronomy and Space Science Institute, 776, Daedeokdae-ro, Yuseong-gu, Daejeon 34055, Republic of Korea}

\author{Arman Shafieloo\orcidlink{0000-0001-6815-0337}}
\email{shafieloo@kasi.re.kr}
\affiliation{Korea Astronomy and Space Science Institute, 776, Daedeokdae-ro, Yuseong-gu, Daejeon 34055, Republic of Korea}
\affiliation{University of Science and Technology, 217 Gajeong-ro, Yuseong-gu, Daejeon 34113, Republic of Korea}

\begin{abstract}

We revisit the physics of the acoustic peaks and illustrate how the early Universe information in the cosmic microwave background (CMB) can be disentangled from the late-time signatures imprinted by gravitational lensing and the integrated Sachs--Wolfe (ISW) effect. Applying this separation to Planck PR4, ACT DR6 and SPT-3G D1 temperature and polarization data combined with the joint lensing reconstruction, we update the model-independent constraints on the angular acoustic scale and the physical densities, $100\theta_*=1.04100\pm0.00021$, $\omega_{\rm b}=(2.2440\pm0.0099)\times10^{-2}$ and $\omega_{\rm bc}=0.1410\pm0.0011$. These constraints are consistent with the previous Planck PR4-only analysis and remain stable for extended late-time cosmologies beyond $\Lambda$CDM.
Adding ACT DR6 to the combination favors a higher $\omega_{\rm b}$ and, through their correlation, a higher primordial tilt $n_{\rm s}$ than PR4 alone.
The reconstructed lensing spectrum exceeds that of $\Lambda$CDM, with two features near $L\simeq50$ and $L\sim10^3$, and the excess is reduced when WMAP replaces the Planck primary spectra.
Marginalizing over the lensing-induced peak smoothing, rather than tying it to the late-time model, lowers the preference of the CMB and DESI DR2 baryon acoustic oscillations (BAO) for $w_0w_a$CDM over $\Lambda$CDM from $3.0\sigma$ to $1.9\sigma$, or from $3.2\sigma$ to $2.4\sigma$ when Union3.1 supernovae are added.
We further use the acoustic prior to predict BAO distances from CMB data alone under several late-time expansion histories, testing their consistency with DESI measurements assuming standard pre-recombination physics.
\end{abstract}

\maketitle
\end{CJK*}

\section{Introduction}\label{sec:intro}

The cosmic microwave background is often summarized as a measurement of the early Universe\footnote{The distinction between the early and late Universe is a steadily evolving target, once roughly synonymous with the nearby Universe spanning a few Mpc, now routinely extending to redshifts of a few.}. This is largely true with acoustic oscillations, baryon loading, radiation driving, neutrino phase shifts, and the photon-diffusion damping tail are all determined by (pre-)recombination physics. The photons, emitted from the last-scattering shell and measured in our detectors, however, traverse the Universe and are modified by low-redshift physics through the late integrated Sachs--Wolfe (ISW) effect~\citep{Sachs1967}, reionization, gravitational lensing, and astrophysical foregrounds (galactic dust, point sources, cosmic infrared background, and Sunyaev--Zel'dovich secondary anisotropies~\citep{Zeldovich:1969,Sunyaev:1972,Sunyaev:1980}). Despite these modifications, the CMB, through its acoustic features, provides one of the most precise standard rulers in cosmology.

A standard fit to $\Lambda$CDM or its proposed extensions using the CMB therefore provides two complementary kinds of information. The primary unlensed spectra constrain pre-recombination physics, while the specified late-time expansion and growth model determines how those spectra are projected, lensed, and interpreted. To test late-time physics, we would prefer a conservative prior that carries only the former while remaining as agnostic as possible about the latter. A prior derived from the full fit inherits assumptions on the expansion history, curvature, a lensing prediction, or a treatment of perturbations that one intends to test. Thus, we would like to retain only the information genuinely supplied by the acoustic physics, and marginalize over or remove the late-time ingredients that would otherwise enter the prior as implicit assumptions.

The way out is to compress the CMB into a few early Universe parameters that calibrate the acoustic physics while leaving the late-time expansion free~\citep{Wang:2007mza,Mukherjee:2008kd}. This approach was introduced in Ref.~\cite{Vonlanthen:2010cd} and developed in Refs.~\cite{Audren:2012wb,Audren:2013nwa}, and most recently Ref.~\cite{Lemos:2023qoy} updated these early Universe priors with Planck PR4 data. We follow this approach closely, update these acoustic priors with the latest CMB data combination, and discuss its ramifications for current discourse on the dynamical dark energy.

The remainder of the paper constructs the compressed prior from the physics of the primary spectra and then applies it to late-time inference. \Cref{sec:physics} develops the projection picture, shows how the angular scales lead to the compressed prior, and identifies the late-time effects that must be cut, marginalized, or measured externally. \Cref{sec:data} describes the CMB data combination and the cosmological inference pipeline. \Cref{sec:results} presents the updated constraints, tests the mutual consistency of the CMB data sets and the robustness to likelihood choices, checks that the three parameters relevant to late times are stable across extensions of $\Lambda$CDM, and provides the resulting prior for direct use. \Cref{sec:lensing} tests the consistency of the reconstructed lensing with specific late-time models. \Cref{sec:desi} %
applies the compressed prior with DESI
to disentangle the dynamical dark energy preference from
the CMB lensing excess, and then tests the prior's predictions for the DESI distances tracer by tracer. We discuss our results in the context of recent compressions and conclude in \Cref{sec:conclusions}. The appendices compare the acoustic-scale conventions of CAMB and CLASS (\Cref{app:camb_class}) and test the dependence on the low-$\ell$ EE likelihood (\Cref{app:lowell_ee}) and on the lensing-spline prior (\Cref{app:lensing_spline_priors}).

\section{CMB Physics and Late-Time Agnostic Compression}
\label{sec:physics}

In the early Universe the comoving mean free path of the photons stays small, $\mathcal{O}(1)\Mpc$, nearly two orders of magnitude below the comoving sound horizon at recombination. On scales well above the mean free path, Thomson scattering keeps the photons tightly coupled to the free electrons, and Coulomb scattering in turn binds those electrons to the baryons, so the three components share a common bulk velocity and move as a single fluid. The density perturbations in this fluid undergo acoustic oscillations in which gravitational infall compresses an overdense region and the photon pressure offers the countering force.

As the Universe expands and cools down, free electrons and protons combine into neutral atoms. This `recombination' proceeds gradually over an extended range of redshift rather than at a single instant. As the free-electron fraction falls, the Thomson scattering rate drops sharply, the photon mean free path grows, and the photons decouple from the baryons and begin to free-stream. We define the last-scattering redshift $z_*$ by
\begin{equation}
 \tau_{\rm rec}(z_*)=1,
 \label{eq:zstar}
\end{equation}
where $\tau_{\rm rec}$ is the optical depth associated with recombination. The probability density for a photon to undergo its final scattering at conformal time $\eta$ is given by the visibility function $g(\eta)=\dot{\tau}e^{-\tau}$, which becomes sharply peaked near decoupling. Here $\dot{\tau}=n_e\sigma_T a$ is the differential optical depth to Thomson scattering. For the standard recombination history, $z_*\simeq1090$ lies close to the maximum of $g(\eta)$, whose half width at half maximum of roughly $\Delta z_*\sim80\text{--}100$ corresponds to $\Delta\eta_*\simeq20\Mpc$.

The decoupling of plasma constituents freezes the acoustic pattern, so the phase of each Fourier mode is set by how far a sound wave could travel before photon decoupling. This comoving sound horizon is expressed as
\begin{align}
 r_s(z_*)&=\int_{z_*}^{\infty}\frac{c_s(z)}{H(z)}\,dz,
 \label{eq:rs}\\
 c_s^2(z)&=\frac{c^2}{3[1+R(z)]},
 \label{eq:cs}
\end{align}
where $R=3\rho_b/(4\rho_\gamma)$ measures the baryon inertia relative to the photon pressure. Initial adiabatic conditions sourced during inflation give the Fourier modes a coherent temporal phase, so modes caught at successive extrema at recombination satisfy
\begin{equation}
 k_m r_s(z_*)\simeq m\pi
 \label{eq:acoustic_harmonics}
\end{equation}
for $m=1,2,\ldots$. As the coupling weakens toward decoupling, the photons develop a local radiation quadrupole that sources linear polarization, so the same physics produces the sequence of acoustic peaks in both the temperature and the polarization spectra, although the latter is offset by a $\pi/2$ phase shift because the polarization is sourced by the fluid velocity~\citep{Hu:1996mn,HuDodelson:2002,Hu:2008cmb}.

Converting these wavenumbers ($k_m$) into an angle on the sky requires the distance to the last-scattering shell, which for a spatially flat universe is the transverse comoving distance\footnote{For $\Omega_K\neq0$, the integral in \Cref{eq:DM} gives the radial comoving distance $\chi(z_*)$, and the transverse distance follows as $D_M=(c/H_0\sqrt{|\Omega_K|})\sinh(\sqrt{|\Omega_K|}H_0\chi/c)$ for an open geometry, with $\sinh$ replaced by $\sin$ for a closed one.}
\begin{equation}
 D_M(z_*)=c\int_0^{z_*}\frac{dz}{H(z)}.
 \label{eq:DM}
\end{equation}

The projection from wavenumber to multipole follows from the line-of-sight solution. The visibility width is small compared with $D_M(z_*)\simeq14\Gpc$, so $\Delta\eta_*/D_M(z_*)\ll1$, which allows the sources to be evaluated on the shell and projected through nearly the same distance~\citep{Hu:2008cmb,HuDodelson:2002}.
Suppressing the Doppler, polarization, and integrated-source terms, the temperature transfer function reduces to
\begin{equation}
 \Theta_\ell(k)\simeq[\Theta+\Psi](k,\eta_*)\,j_\ell\!\left(kD_M(z_*)\right),
 \label{eq:los_shell}
\end{equation}
and the observed spectrum follows as $C_\ell=4\pi\int d\ln k\,\mathcal{P}_\zeta(k)\,|\Theta_\ell(k)/\zeta(k)|^2$ for a primordial curvature power spectrum $\mathcal{P}_\zeta$. The source $[\Theta+\Psi]$ carries the intrinsic temperature and gravitational-redshift pattern imprinted before recombination. All of the $\ell$ dependence sits in the spherical Bessel function, which for $\ell\gg1$ peaks near
\begin{equation}
\ell\simeq kD_M(z_*).
\label{eq:ell_k_map}
\end{equation}

The sound horizon subtends the angle
\begin{equation}
 \theta_*=\frac{r_s(z_*)}{D_M(z_*)},
 \label{eq:acoustic_angle}
\end{equation}
which appears in the spectra at the multipole $\ell_A=\pi/\theta_*$. Under the same projection, the harmonic series of \Cref{eq:acoustic_harmonics} places the peaks at $\ell_m\simeq(m-\varphi_m)\ell_A$, so their near-uniform spacing $\Delta\ell\simeq\ell_A$ measures $\theta_*$ to high precision, with $\varphi_m$ accounting for phase shifts from potential evolution and free-streaming radiation~\citep{Hu:1996mn,Bashinsky:2003tk,Follin:2015hya,Ge:2022qws}.

\Cref{eq:los_shell} separates the two halves of the problem. Pre-recombination physics fixes the source as a function of $k$, and the late-time expansion enters only through the argument of $j_\ell$. Changing the low-redshift expansion history at fixed early physics therefore rescales the multipole axis through $D_M(z_*)$ and leaves the pattern intact.

While the peak spacing measures $\theta_*$, the peak heights carry information to determine the physical densities that fix $r_s$ in absolute units. Baryons add inertia to the oscillating fluid without contributing pressure, so a larger baryon loading $R(z_*)$ both lowers the sound speed in \Cref{eq:cs} and, acting through the gravitational forcing, displaces the equilibrium of the oscillation toward compression. Compressions into the potential wells are then deeper than the intervening rarefactions, which raises the odd acoustic peaks relative to even peaks. Because the measured CMB temperature fixes the photon density, the amplitude of this odd--even modulation measures $\omega_{\rm b}$, which in turn sets the baryonic contribution to the sound horizon.

The matter density is measured through the response of the oscillator to matter--radiation equality. A mode that enters the horizon during radiation domination has its gravitational potential sourced by the photon--baryon fluid itself. Photon pressure stops the fluid from growing fast enough to sustain that potential, so the potential decays away over the first compression. The fluid rebounds with the restoring gravity already gone, and the oscillation amplitude is boosted. Modes entering after equality miss this radiation driving, so a larger $\omega_{\rm bc}$\footnote{We define $\omega_{\rm bc}=\omega_{\rm b}+\omega_{\rm cdm}$ for baryons and cold dark matter, while the total matter density adds the present-day massive-neutrino contribution, so that $m=b+{\rm cdm}+\nu$. For the standard light-neutrino sector, the neutrinos remain relativistic at matter--radiation equality and are therefore included in $\omega_r$. The equality redshift is obtained from the redshift-dependent relativistic and nonrelativistic energy densities rather than the approximate ratio in \Cref{eq:equality}.} moves equality earlier, leaves fewer modes boosted, and lowers the acoustic peaks relative to the large-angle Sachs--Wolfe plateau. The suppression is partly offset because weaker driving leaves deeper potentials at $z_*$ and so strengthens the odd--even modulation~\citep{Hu:2000ti}. The conventional equality scale then satisfies
\begin{align}
 k_{\rm eq}=\frac{a_{\rm eq}H_{\rm eq}}{c}, \qquad
 1+z_{\rm eq}\simeq\frac{\omega_{\rm bc}}{\omega_r}.
 \label{eq:equality}
\end{align}
Its observable imprint is a broad change in the peak-height envelope, including the early ISW contribution, rather than a localized feature at a single multipole. Once the radiation content, neutrino sector, and primordial spectrum are specified, the scale dependence of this driving constrains $\omega_{\rm bc}$. The second and third peaks help distinguish the matter-density response from baryon loading because the two effects alter their relative heights differently~\citep{HuDodelson:2002,Hu:2008cmb}.

Photon diffusion tests the same calibration through different microphysics. The finite photon mean free path generates shear viscosity and heat conduction, which dissipate perturbations below the diffusion length via Silk damping. With $d\eta=c\,dt/a$, the wavenumber associated with such damping is given by
\begin{equation}
 k_d^{-2}(z_*)=\int_0^{\eta_*}
 \frac{d\eta}{a\sigma_T n_e}
 \frac{R^2+\frac{16}{15}(1+R)}{6(1+R)^2}.
 \label{eq:diff_distance}
\end{equation}
The transfer amplitude is suppressed approximately as $\exp[-(k/k_d)^2]$, and the finite duration of recombination adds a similar Landau damping that further smooths the projected cutoff. For an approximately uniform rescaling of the early expansion rate at fixed ionization history, $r_s\propto H^{-1}$ while $k_d^{-1}\propto H^{-1/2}$. The damping tail therefore tests the expansion rate and recombination history inferred from the lower peaks, with additional dependence on $Y_P$ and $N_{\rm eff}$~\citep{Hu:1996vq,Hu:2008cmb,Hou:2011ec}.

Equality and diffusion introduce the further angular markers $\ell_{\rm eq}=k_{\rm eq}D_M(z_*)$ and $\ell_d=k_dD_M(z_*)$, which share the projection distance $D_M(z_*)$ with the acoustic scale $\ell_A$. This common distance cancels from the ratios
\begin{align}
 \frac{\ell_{\rm eq}}{\ell_A}
 &=\frac{k_{\rm eq}r_s(z_*)}{\pi},
 \nonumber\\
 \frac{\ell_d}{\ell_A}
 &=\frac{k_dr_s(z_*)}{\pi},
 \label{eq:scale_ratios}
\end{align}
which are fixed by the pre-recombination physics alone. The three markers nonetheless enter the spectra in distinct ways. The acoustic scale is measured from the periodic sequence of peak positions and supplies $\theta_*$, whereas equality appears as a broad transition in the driving envelope and diffusion sets the high-$\ell$ damping scale. 
The full likelihood fits the peak locations, relative heights, phases, and damping jointly rather than treating the three markers as independent measurements or assigning one parameter to each scale.

In addition to pre-recombination physics, several further contributions modify the anisotropies or contaminate the spectra observed today. Each of these carries late-time information that would enter the prior as an implicit assumption, so we remove, marginalize, or externally constrain each of them, following the early-$\Lambda$CDM compression method of Ref.~\cite{Lemos:2023qoy}. 

\begin{itemize}
 \item \textit{Late ISW:} The late ISW effect arises when the expansion departs from matter domination, as dark energy or curvature becomes dynamically important at low redshift and the gravitational potentials decay along the line of sight. Its main contribution lies at large angular scales ($\ell<30$), where cosmic variance also limits the available information. We remove the dominant sensitivity by excluding TT at $\ell<30$, set the ISW term of the temperature source to zero for $z<15$ at the remaining multipoles, and add a residual template with fixed shape and free amplitude,
\begin{align}
 C_\ell^{TT}\longrightarrow
 C_\ell^{TT,\,{\rm w/o}\,\lisw}+&A_{\lisw}T_\ell^{TT,\lisw},
 \nonumber\\
 \frac{\ell(\ell+1)}{2\pi}T_\ell^{TT,\lisw}&=\left(\frac{30}{\ell}\right)^{2}\mu{\rm K}^2,\quad \ell\geq30,
 \nonumber\\
 A_{\lisw}&\in[0,100].
 \label{eq:isw_template}
\end{align}

\item \textit{CMB lensing:} The unlensed primary spectra are determined by the primordial fluctuations and recombination transfer functions. Intervening structure then remaps the resulting primary anisotropies through deflections by a lensing potential $\phi$ sourced mostly at $z\sim1$--3. CMB lensing cannot be removed by a multipole cut because it smooths the acoustic peaks and transfers power throughout the small-scale spectra. 
The same remapping produces off-diagonal correlations between otherwise independent multipoles, which allow $C_L^{\phi\phi}$ to be reconstructed directly with quadratic estimators~\citep{HuOkamoto:2002,Okamoto:2003zw}. A theoretical prediction for $C_L^{\phi\phi}$ would import assumptions about late-time structure growth, and a single $A_L$ rescaling would still retain that prediction's shape even where Planck peak smoothing and the direct reconstruction show mild shape-dependent tension~\citep{Motloch:2018pjy,Motloch:2019gux}. We instead use a smooth log-amplitude spline (defined in Section~\ref{sec:data}), constrained by the lensing reconstruction and applied to lens the primary spectra at every sampled point, so that the reconstruction data determine scale-dependent departures directly. Marginalizing the spline amplitudes retains the measured smoothing without importing a specified late-time growth model.

 \item \textit{Reionization:} Reionization rescatters a fraction of the CMB photons after the first neutral atoms have formed. On scales well inside the horizon at reionization, the dominant effect on TT, TE, and EE is a suppression of primary power by $e^{-2\tau}$, so the high-$\ell$ spectra measure the amplitude combination $A_\mathrm{s}e^{-2\tau}$. Low-$\ell$ EE polarization supplies a separate constraint on $\tau$ through the quadrupole rescattered during reionization, but its interpretation depends on the assumed ionization history. We retain $\tau$ as a parameter and marginalize to prevent the acoustic calibration from relying on a fixed reionization history~\citep{Planck-2018-likelihoods,Planck-2018-cosmology,Lemos:2023qoy}.

 \item \textit{Astrophysical foregrounds:} Emissions from Galactic dust, radio and infrared sources, the cosmic infrared background, and secondary anisotropies such as thermal and kinetic Sunyaev--Zel'dovich signals act as foreground contaminants for the primary CMB. Their frequency and angular dependence allow most of this emission to be separated from the blackbody CMB signal, but residual foreground and calibration uncertainties remain in the power spectra. For the sake of computational convenience, we use the foreground-marginalized lite likelihoods, so these nuisance directions contribute to the uncertainty without being absorbed into the inferred acoustic parameters.
\end{itemize}

The acoustic scale and physical-density calibration retained from the primary spectra are summarized by the marginalized posterior for
\begin{equation}
 \left(\theta_*,\,\omega_{\rm b},\,\omega_{\rm bc}\right).
 \label{eq:early_vector}
\end{equation}
The construction assumes adiabatic initial conditions with a power-law spectrum, and a standard recombination and neutrino sector.
We marginalize over the primordial tilt, the amplitude combination $A_\mathrm{s}e^{-2\tau}$, the residual ISW amplitude, the lensing spline, and the foreground nuisance parameters.

The physical densities calibrate $r_s(z_*)$ in absolute units, so the measured angular size of the sound horizon at last scattering, $\theta_*$, determines the transverse comoving distance $D_M(z_*)$. With the late-time contributions removed or marginalized, the primary acoustic spectra cannot by themselves break the geometric degeneracy that prevents a unique determination of $H_0$~\citep{Eisenstein:1998tu,Efstathiou:1998xx}, because different low-redshift expansion histories can yield the same $D_M(z_*)$~\citep{Hu:2004kn,Linder:2007ka}. The tight $H_0$ constraint commonly quoted from the CMB therefore arises only after restricting the late-time background model (such as $\Lambda$CDM) or adding external distance information that breaks the degeneracy between $H_0$ and an evolving dark energy sector~\citep{Hu:2008cmb}.
BAO measure $D_M(z)$ and $D_H(z)=c/H(z)$ at low redshift in units of $r_d$, the sound horizon at the end of the Compton drag epoch~\citep{1998ApJ...496..605E}. Because the baryons are released slightly after the photons decouple, $r_d\simeq1.02\,r_s(z_*)$, so the physical densities that calibrate $r_s(z_*)$ calibrate $r_d$ as well. The offset depends on the baryon loading and the ionization history~\citep{Hu:1995en} but barely changes across extensions~\citep{Pogosian:2024ykm,Mirpoorian2025}.

\section{Data}\label{sec:data}

In this work, we use an updated combination of CMB measurements, including Planck PR4 (NPIPE)~\cite{Planck-2020-NPIPE}, ACT DR6~\cite{AtacamaCosmologyTelescope:2025blo}, and SPT-3G D1~\cite{SPT-3G:2025bzu}. For Planck-only analyses, we adopt the CamSpec likelihood~\cite{Rosenberg:2022sdy}. When Planck PR4 is combined with ACT or SPT, we instead use the CamSpec-NPIPE-lite likelihood with foreground nuisances marginalized, following the recommendation of~\cite{Jense:2025wyg}. We also use the corresponding lite likelihoods for SPT~\cite{SPT-3G:2025bzu,SPT-3G:2026krg,Balkenhol:2024sbv} and ACT~\cite{AtacamaCosmologyTelescope:2025blo}, so that foreground and calibration nuisance parameters are consistently marginalized in the combined analyses. When combining SPT with Planck, we follow \citet{SPT-3G:2025bzu} and do not impose multipole cuts, because the sky overlap between the two surveys is small. For ACT, we avoid double-counting strongly correlated modes by removing the small-scale Planck information and the large-scale ACT information. The transition multipoles are broadly chosen according to the relative signal-to-noise of the two experiments and are set at $\ell=1500$ in TT, $\ell=1000$ in TE, and $\ell=600$ in EE. For all data-set combinations, we include the Planck SRoll2 low-$\ell$ EE likelihood~\cite{Delouis:2019bub}. In Appendix~\ref{app:lowell_ee}, we show that removing this likelihood has a negligible effect on the five-parameter early Universe constraints used in our compression.

For CMB lensing, we fit the \texttt{actplanckspt3g\_baseline} variant of the joint likelihood of Ref.~\cite{ACT:2025qjh}\footnote{\url{https://github.com/qujia7/spt_act_likelihood}} together with the primary CMB spectra, with \texttt{lens\_only} set to \texttt{False}. This likelihood combines the Planck PR4~\cite{Carron:2022eyg} and ACT DR6~\cite{ACT:2023ubw,ACT:2023dou,ACT:2023kun} lensing reconstructions with the SPT-3G lensing bandpowers measured with the MUSE estimator~\cite{SPT-3G:2024atg}, including their cross-covariance, and we adopt the baseline multipole range for ACT. We refer to the combination of the PR4, ACT DR6, and SPT-3G D1 primary spectra with the multipole cuts above, the SRoll2 low-$\ell$ EE likelihood, and combined lensing likelihoods as the baseline dataset, denoted PR4+ACT DR6+SPT-3G D1. For analyses based on Planck data alone, the spline for $C_L^{\phi\phi}$ uses five nodes. When ACT or SPT data are included, we use six nodes in order to cover the additional small-scale information provided by these experiments. The node positions are chosen following~\cite{Lemos:2023qoy}, namely $L = \{7, 44, 125, 600, 1600, 3100\}$. We impose a broad flat prior on each node amplitude, $\log\left([L(L+1)]^2C_L^{\phi\phi}/2\pi\right) \in [-22, -14]$, which is sufficiently wide to avoid imposing a restrictive late-time model prior on the lensing reconstruction. In Appendix~\ref{app:lensing_spline_priors}, we verify that increasing the number of nodes and changing the node-prior parameterization leave the early Universe constraints essentially unchanged.

In \cref{sec:desi}, we combine the CMB with DESI DR2 BAO~\cite{DESI:2025zgx} and with the Union3.1 compilation of type Ia supernovae~\cite{Hoyt:2026fve,Rubin:2026qdt}, analyzed with the UNITY1.8 model. For \cref{tab:w0wa_significance}, we follow Refs.~\cite{DESI:2025zgx,Lodha2025} and quote $\Delta\chi^2_{\rm MAP}$, the decrease in twice the negative log posterior at its maximum from $\Lambda$CDM to $w_0w_a$CDM, after removing the volume of the flat priors $w_0\in[-3,3]$ and $w_a\in[-10,10]$, with $w_0+w_a\le0$ so that dark energy grows no faster than matter toward early times. We convert it to a two-sided Gaussian significance for two extra parameters using Wilks' theorem.

The theoretical spectra are computed with a version of CAMB~\cite{Lewis:1999bs} modified to implement the late-ISW treatment and the lensing spline as mentioned in \Cref{sec:physics}. We sample $\ln(10^{10}A_\mathrm{s})$, $n_\mathrm{s}$, $100\theta_{\rm MC}$, $\omega_{\rm b}$, $\omega_{\rm c}$, and $\tau$ with broad uniform priors, together with the residual ISW amplitude, the six spline amplitudes, and the calibration and foreground nuisance parameters of each likelihood. The acoustic scale $100\theta_*$ and $\omega_{\rm bc}=\omega_{\rm b}+\omega_{\rm c}$ are derived parameters. We use  CosmoRec~\cite{Chluba:2010fy} for the recombination, fix $N_{\rm eff}=3.044$ and assume a single massive neutrino with $\sum m_\nu=0.06\,{\rm eV}$. Posterior sampling is performed with Cobaya~\cite{Torrado:2020dgo}. We run the Markov chains until they satisfy the Gelman--Rubin convergence criterion~\cite{GelmanRubin}, requiring $R-1<0.01$ for all sampled parameters. For the best-fit points, we start from the maximum a posteriori (MAP) points of the chains and refine them with the iminuit minimizer~\cite{iminuit}.

\section{Results}\label{sec:results}
\begin{figure*}[htp!]
 \centering
 \includegraphics[width=0.9\textwidth]{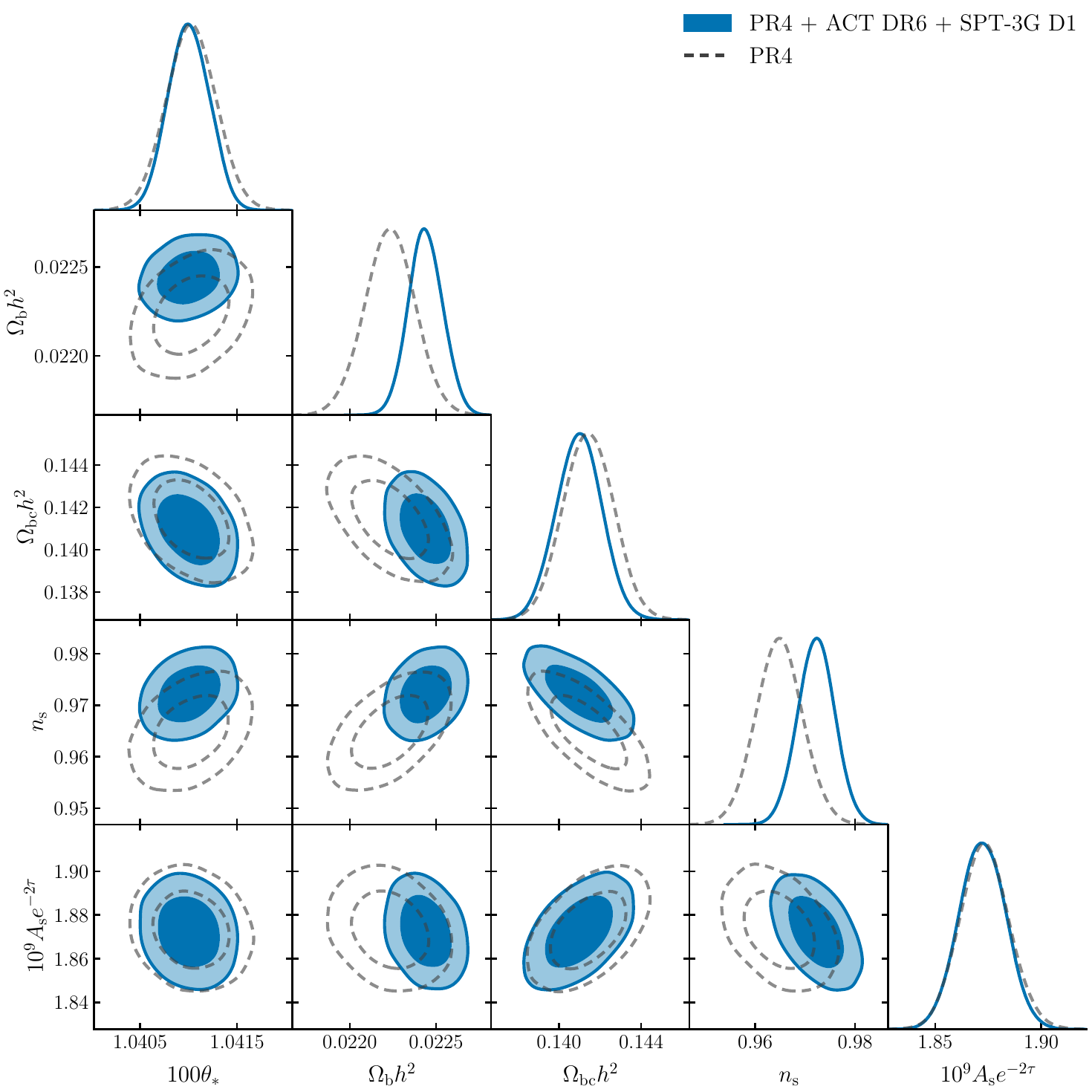}
 \caption{68\% and 95\% credible regions of the early-$\Lambda$CDM parameters from the baseline Planck PR4+ACT DR6+SPT-3G D1 (filled) and from PR4 alone~\cite{Lemos:2023qoy} (dashed). Adding ACT and SPT tightens $\omega_{\rm b}$ most and shifts $\omega_{\rm b}$ and $n_\mathrm{s}$ toward higher values.}
 \label{fig:baseline}
\end{figure*}

Having specified the compression and the data, we now turn to the constraints on the early-$\Lambda$CDM parameters from combining the space-based Planck PR4 with the ground-based ACT DR6 and SPT-3G D1 data. We first compare these constraints with the PR4-only results of \citet{Lemos:2023qoy} in Fig.~\ref{fig:baseline}\footnote{%
Note that these PR4-only results use the SimAll low-$\ell$ EE likelihood~\cite{Planck-2018-likelihoods} and Gaussian priors $\omega_{\rm b}\sim\mathcal{N}(0.0222,0.0005)$ and $n_\mathrm{s}\sim\mathcal{N}(0.96,0.02)$, whereas our combined data set uses SRoll2~\cite{Delouis:2019bub} and flat priors. Here we use the publicly available results from \url{https://github.com/cmbant/PlanckEarlyLCDM}. We note that the prior on the final lensing node in those results may have been inadvertently too narrow,
but we find its impact to be negligible.}, and then test whether they depend on the choice of data set or on the assumed late-time cosmology.

Besides the three compressed parameters, we report the primordial tilt $n_\mathrm{s}$ and the overall scaling amplitude $A_\mathrm{s}e^{-2\tau}$.%
With ACT DR6 and SPT-3G D1 added, the acoustic scale remains fairly stable as expected, with its uncertainty shrinking by 18\%. The baryon density gains the most, with a 32\% smaller error and an upward shift of $1.4\sigma$ in units of the PR4-only error, whereas $\omega_{\rm bc}$ moves by only $0.4\sigma$ and tightens by 9\%. %
The spectral index $n_\mathrm{s}$ shifts upward together with $\omega_{\rm b}$, as the $\omega_{\rm b}$--$n_\mathrm{s}$ panel of \cref{fig:baseline} shows.

Table~\ref{tab:compressed} summarizes these constraints alongside the individual ground-based data sets and the baseline with WMAP replacing the Planck large scale multipoles.
To quantify these shifts, we use the squared Mahalanobis distance
\begin{equation}
 D_{AB}^2 =
 \left(\bar{\boldsymbol{\theta}}_A-\bar{\boldsymbol{\theta}}_B\right)^T
 \left(C_A+C_B\right)^{-1}
 \left(\bar{\boldsymbol{\theta}}_A-\bar{\boldsymbol{\theta}}_B\right),
 \label{eq:mahalanobis_distance}
\end{equation}
where $C_{A,B}$ are the posterior covariance matrices for data sets $A$ and $B$, and $\bar{\boldsymbol{\theta}}_{A,B}$ are the corresponding posterior means. Assuming approximately Gaussian posteriors, we convert $D^2_{AB}$ into a $p$-value with as many degrees of freedom as compared parameters and quote the equivalent two-sided Gaussian significance.
The squared Mahalanobis distance between PR4-only and PR4+ACT DR6+SPT-3G D1 corresponds to an equivalent two-sided Gaussian significance of $0.44\sigma$, both for the full set of five parameters and for the first three parameters relevant to the late Universe. The baseline shares the Planck spectra below the ACT cuts with PR4, so the two posteriors are correlated and $C_A+C_B$ overestimates the scatter of their difference, whereas the update difference in means, appropriate when one data set extends the other, has covariance $C_A-C_B$~\cite{Raveri:2018wln}. With this covariance, $\omega_{\rm b}$ shifts by $1.9\sigma$ and $\omega_{\rm bc}$ by $0.9\sigma$, while $\theta_*$ barely moves, and the largest shift, $2.6\sigma$, is in $n_\mathrm{s}$. Since the baseline also drops the Planck spectra above the ACT cuts and uses a different low-$\ell$ EE likelihood and priors, it is not a strict extension of PR4, and the true significance likely lies between the two estimates.
The upward shift in $n_\mathrm{s}$ has the same direction as in analyses that assume $\Lambda$CDM for both the early and late Universe~\cite{AtacamaCosmologyTelescope:2025blo,SPT-3G:2025bzu}. As in the P-ACT combination~\cite{AtacamaCosmologyTelescope:2025blo}, it comes with a rise in $\omega_{\rm b}$, because the larger-scale peaks correlate the two parameters positively while the damping tail anticorrelates them, so the combined data break the degeneracy. Since our compression marginalizes over the late-time lensing and ISW, this shift arises from the primary spectra, and unless it is caused by systematic effects it points to early Universe physics, including inflation, rather than to an inaccurate description of the late Universe.

\begin{table*}[htp!]
  \centering
  \caption{Posterior means and standard deviations of the early-$\Lambda$CDM parameters in the compression for each CMB data combination, all including the SRoll2 low-$\ell$ EE likelihood and CMB lensing reconstruction. The PR4 column is our own analysis, distinct from the public chains of Ref.~\cite{Lemos:2023qoy} in Fig.~\ref{fig:baseline}.}
  \label{tab:compressed}
  \begin{tabular}{l@{\hspace{24pt}}>{\bfseries\boldmath}ccccc}
    \toprule
    Parameter & PR4 + ACT + SPT & PR4 & ACT DR6 & SPT-3G D1 & WMAP + ACT + SPT \\
    \midrule
    $100\theta_*$ & $1.04100 \pm 0.00021$ & $1.04107 \pm 0.00026$ & $1.04086 \pm 0.00035$ & $1.04087 \pm 0.00068$ & $1.04095 \pm 0.00029$ \\
    $\omega_{\rm b}$ & $0.022440 \pm 0.000099$ & $0.02226 \pm 0.00016$ & $0.02259 \pm 0.00017$ & $0.02222 \pm 0.00021$ & $0.02249 \pm 0.00011$ \\
    $\omega_{\rm bc}$ & $0.1410 \pm 0.0011$ & $0.1412 \pm 0.0013$ & $0.1456 \pm 0.0032$ & $0.1453 \pm 0.0053$ & $0.1434 \pm 0.0022$ \\
    $n_\mathrm{s}$ & $0.9722 \pm 0.0037$ & $0.9658 \pm 0.0055$ & $0.9661 \pm 0.0090$ & $0.942 \pm 0.019$ & $0.9657 \pm 0.0051$ \\
    $10^9 A_\mathrm{s} e^{-2\tau}$ & $1.872 \pm 0.011$ & $1.870 \pm 0.012$ & $1.886 \pm 0.020$ & $1.933 \pm 0.037$ & $1.889 \pm 0.012$ \\
    \bottomrule
  \end{tabular}
\end{table*}

\begin{figure*}[htp!]
 \centering
 \includegraphics[width=0.9\textwidth]{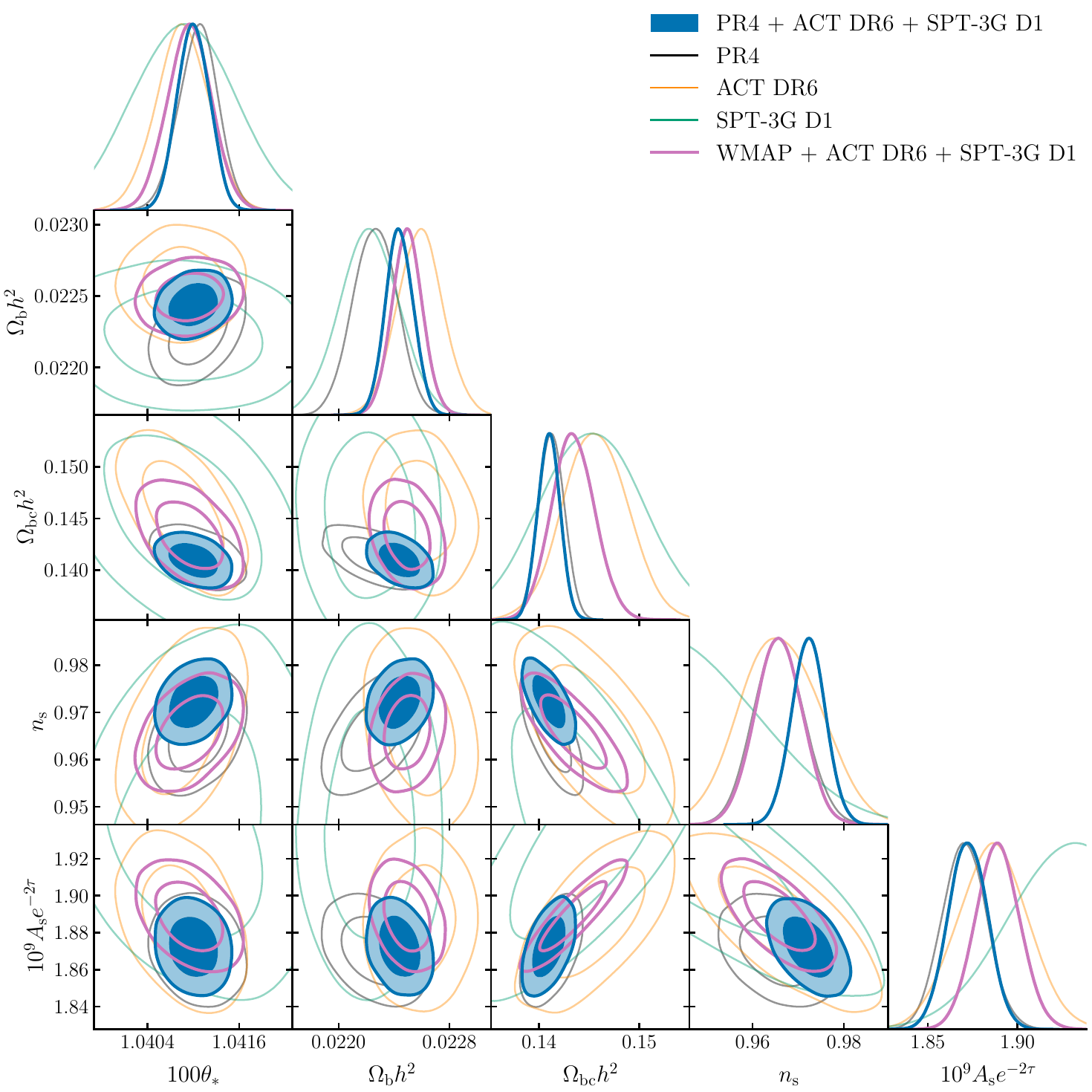}
 \caption{Early-$\Lambda$CDM parameters from PR4, ACT DR6, and SPT-3G D1 separately, from their baseline combination (filled), and from WMAP+ACT DR6+SPT-3G D1, in which the WMAP spectra replace the Planck spectra below the ACT multipole cuts and the lensing reconstruction is unchanged. The three experiments are mutually consistent.}
 \label{fig:PASW}
\end{figure*}

\begin{table}[htp!]
  \centering
  \caption{Consistency between CMB data sets, quoted as the $p$-value with the
  equivalent two-sided Gaussian significance in parentheses. The 3D column uses
  the compressed subspace $\{\theta_*,\omega_{\rm b},\omega_{\rm bc}\}$,
  and the 5D column the full early-$\Lambda$CDM parameter space
  $\{\theta_*,\omega_{\rm b},\omega_{\rm bc},n_\mathrm{s},A_\mathrm{s}e^{-2\tau}\}$.}
  \label{tab:dataset_consistency}
  \begin{tabular}{@{}lc@{\hspace{18pt}}c@{}}
    \toprule
    Datasets & 3D & 5D \\
    \midrule
    PR4 vs ACT DR6 & 0.180 ($1.34\sigma$) & 0.218 ($1.23\sigma$) \\
    PR4 vs SPT-3G D1 & 0.905 ($0.12\sigma$) & 0.691 ($0.40\sigma$) \\
    ACT DR6 vs SPT-3G D1 & 0.596 ($0.53\sigma$) & 0.316 ($1.00\sigma$) \\
    \midrule
    PR4 vs WMAP+ACT+SPT & 0.329 ($0.98\sigma$) & 0.520 ($0.64\sigma$) \\
    \makecell[l]{Baseline vs\\ WMAP+ACT+SPT} & 0.669 ($0.43\sigma$) & 0.856 ($0.18\sigma$) \\
    \bottomrule
  \end{tabular}
\end{table}

The combined constraints assume that the three experiments agree once late-time physics is marginalized. We test this by analyzing PR4, ACT DR6, and SPT-3G D1 separately, each with its full multipole range and its own lensing reconstruction. Since only Planck measures the large-scale polarization well enough to constrain $\tau_\mathrm{reio}$, every analysis includes the SRoll2 low-$\ell$ EE likelihood, and we omit $\tau_\mathrm{reio}$ from the comparison. \Cref{fig:PASW} shows the individual constraints, and \cref{tab:dataset_consistency} lists the resulting significances in the three- and five-dimensional parameter spaces. All pairs agree to within $1.4\sigma$, with the largest difference, $1.34\sigma$ in the compressed space, between PR4 and ACT DR6. These comparisons neglect correlations between the data sets, such as those from the overlapping Planck and ACT sky, which reduce the covariance of the difference, so the quoted significances are slightly underestimated. We have verified the Gaussian approximation for the 3D and 5D posteriors entering these comparisons. For all data combinations, Mardia's multivariate skewness and kurtosis~\cite{Mardia1970} are consistent with a Gaussian within the sampling noise of the chains, and the $68\%$ and $95\%$ regions of the Gaussian approximation enclose the corresponding fractions of the posterior mass.

In the baseline combination, Planck supplies the multipoles below the ACT cuts. In \Cref{fig:PASW}, we also test whether the compressed parameters depend on this choice by replacing the Planck spectra at these multipoles with the WMAP temperature and polarization spectra~\cite{WMAP:2012fli}, combined with ACT DR6, SPT-3G D1, the SRoll2 low-$\ell$ EE likelihood, and the same joint Planck, ACT, and SPT lensing reconstruction as the baseline. The constraints broaden, most visibly in $\omega_{\rm bc}$, and the means move along the degeneracy that links $\omega_{\rm bc}$, $n_\mathrm{s}$, and $A_\mathrm{s}e^{-2\tau}$. The two combinations differ by $0.43\sigma$ in the compressed space and $0.18\sigma$ in the full space. Table~\ref{tab:dataset_consistency} also shows that PR4 alone differs from WMAP+ACT+SPT by $0.98\sigma$ and $0.64\sigma$, so a determination without the Planck spectra at these multipoles agrees with the Planck one.

The baseline also fixes two likelihood choices, the low-$\ell$ EE likelihood and the prior on the lensing spline. Removing SRoll2 weakens the constraint on the optical depth to $\tau_\mathrm{reio}<0.169$ at 95\% confidence, yet it shifts the three compressed parameters by less than $0.06\sigma$ and changes their uncertainties by at most $2.0\%$ (Appendix~\ref{app:lowell_ee}). Increasing the number of spline nodes from six to eleven with wider priors, or splining the ratio to the CAMB lensing spectrum instead, shifts the five early-$\Lambda$CDM parameters by at most $0.09\sigma$ and changes their uncertainties by at most $3.4\%$ (Appendix~\ref{app:lensing_spline_priors}).

The low-$\ell$ EE test also bears on recent proposals that an optical depth above the low-$\ell$ EE value would ease the tension between the CMB and DESI in $\Lambda$CDM and weaken the preference for $w_0w_a$CDM~\cite{Sailer:2025lxj,Jhaveri:2025neg}. \citet{Sailer:2025lxj} attribute the geometric part of this relief to the anticorrelation between the CMB-inferred $\tau$ and $\Omega_m$. Once the lensing spectrum is freed from the model prediction, however, we find that the acoustic parameters calibrating the sound horizon do not track $\tau$, and the anticorrelation must therefore enter through the lensing amplitude and not through the acoustic physics, in agreement with \citet{Jhaveri:2025neg}. %

\begin{figure*}[htb!]
 \centering
 \includegraphics[width=\textwidth]{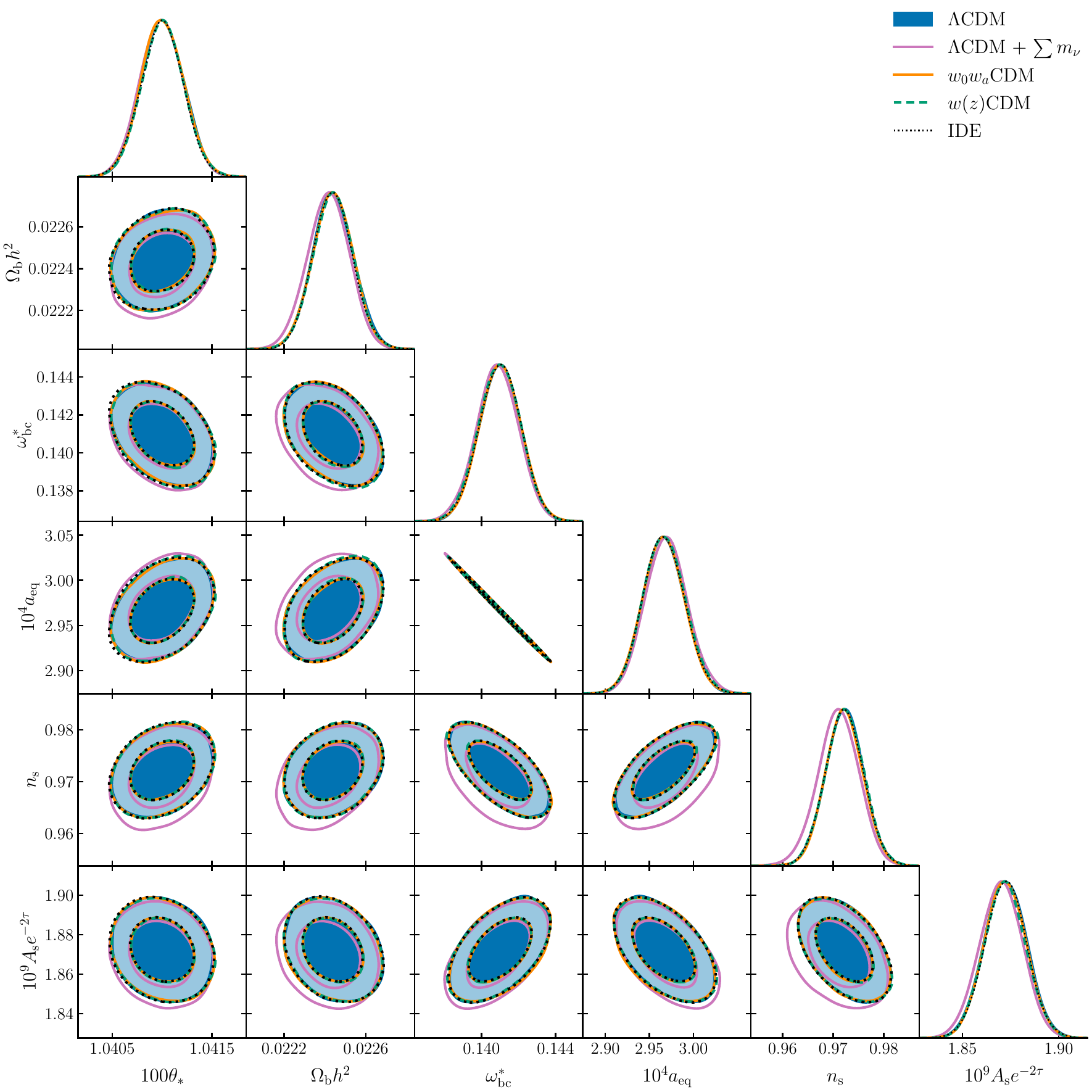}
\caption{Marginalized posterior constraints on the early-$\Lambda$CDM parameters from the baseline data combination for $\Lambda$CDM, $\Lambda$CDM+$\sum m_\nu$, $w_0w_a$CDM, a binned $w(z)$CDM model, and a dark matter--dark energy interaction model, together with the derived scale factor at matter--radiation equality, $10^4a_{\rm eq}$. The compressed parameters and $a_{\rm eq}$ are unchanged across the models, and the largest change, in $n_\mathrm{s}$, comes from the free neutrino mass.}
 \label{fig:late}
\end{figure*}

The tests above vary the data and the likelihoods while keeping the late-time model fixed. The defining requirement of the compressed prior is instead that the inferred early Universe information remain insensitive to the assumed late-time cosmology.
Although construction in \Cref{sec:physics} should already ensure this, nevertheless we verify this expectation by repeating the analysis in $\Lambda$CDM with a free neutrino mass sum, in the $w_0w_a$CDM extension, in a binned $w(z)$CDM model, and in an interacting dark energy (IDE) model~\cite{Amendola:1999er}, and comparing with the baseline. For the neutrino test, we vary the sum of neutrino masses, carried by a single massive species, with a flat prior $\sum m_\nu\in[0,5]$\,eV. The binned model~\cite{Lodha2025} varies $w$ independently in four equally spaced bins over $0<z<2.1$ and in a fifth bin that captures the higher-redshift behavior, with a flat prior $w_i\in[-3,1]$ in each. For the IDE model, we consider an energy transfer between dark energy and cold dark matter of the form
$Q=\beta H\rho_{\rm de}$~\cite{Gavela:2009cy,Salvatelli:2013wra},
where $\beta$ is a dimensionless coupling parameter, with energy flowing from dark energy to dark matter for $\beta>0$, and $\beta=0$ recovers the noninteracting limit. We fix $w=-0.999$ for this test, following Ref.~\cite{Salvatelli:2013wra}, and vary $\beta$ in $[-1,1]$.

When the CDM density departs from $a^{-3}$ after recombination, for example
through an interaction with dark energy or a late-time decay, its present-day
value no longer reflects the density that shapes the acoustic peaks.
We
therefore introduce an effective CDM density $\omega_{\rm cdm}^*$, which for
the IDE model is
\begin{equation}
\omega_{\rm cdm}^{*} = \omega_{\rm cdm} + \beta\,\omega_{{\rm de},0}\frac{a_*^{-3w-\beta}-1}{-3w-\beta} \, ,
\label{eq:omega_cdm_star}
\end{equation}
where $a_*=(1+z_*)^{-1}$, $h=H_0/100$, and
$\omega_{{\rm de},0} = h^2-\omega_{\rm b}-\omega_{\rm cdm}-\omega_\nu . $
Any pre-recombination epoch at which the model asymptotes to $\Lambda$CDM, including a fixed redshift, may be used in place of $a_*$.
We then define the recombination-era baryon and CDM density as
\begin{equation}
\omega_{\rm bc}^{*} \equiv \omega_{\rm b}+\omega_{\rm cdm}^{*}.
\label{eq:omega_bc_star}
\end{equation}
This quantity is relevant for the acoustic peak structure and
provides a common early-time density parameter that can be compared
across the noninteracting and interacting models.

\Cref{fig:late} compares the marginalized constraints on the cosmological parameters obtained in these five late-time scenarios. The posterior distributions are nearly indistinguishable across the models, even though $w_0w_a$CDM and the binned $w(z)$CDM change the late-time expansion and IDE changes the scaling of the matter density at late times. Neither the means nor the uncertainties of $\theta_*$, $\omega_{\rm b}$, and $\omega_{\rm bc}^{*}$ respond to these changes, even in the binned model, which adds five free parameters to the dark energy sector. Had the compressed parameters retained any late-time information, this extra freedom would have broadened them. The early Universe information is thus separated from the late-time model, and one prior serves all of these cosmologies.

As a complementary physical diagnostic, we also consider the scale factor at the matter-radiation equality, $a_{\rm eq}$, such that
\begin{equation}
\rho_{\rm m}(a_{\rm eq})=\rho_{\rm r}(a_{\rm eq}) \, ,
\end{equation} 
computed from the background evolution for each posterior sample, 
where $\rho_{\rm m}$ and $\rho_{\rm r}$ denote the total matter and 
radiation energy densities, respectively. This quantity directly characterizes 
the epoch at which the Universe transitions from radiation to matter domination. 
The inferred epoch of matter-radiation equality is similarly stable, 
with $10^4a_{\rm eq}=2.97\pm0.02$ in all five models.
The stability of $\omega_{\rm bc}^{*}$ and $a_{\rm eq}$ is particularly relevant for the IDE model. 
Although the interaction changes the relation between the present-day CDM density and its value at recombination, 
the baryon and CDM density governing the acoustic peak structure remains tightly constrained. The interaction 
therefore affects the subsequent evolution from recombination to the present day without producing a significant
shift in the early Universe density inferred from the CMB.

Massive neutrinos test a different part of the compression. In a standard CMB analysis, $\sum m_\nu$ is constrained mainly through lensing, since free-streaming suppresses $C_L^{\phi\phi}$ on small scales and with it the smoothing of the acoustic peaks~\cite{Motloch:2019gux}. The spline reconstruction absorbs this effect, and consistent with this we obtain only $\sum m_\nu<0.60$\,eV at 95\% confidence. The largest change is in $n_\mathrm{s}$, which shifts down by $0.3\sigma$ with a 10\% larger uncertainty. The shift comes mainly from masses above $\sim0.3$\,eV, beyond the bound $\sum m_\nu<0.24$\,eV from the full Planck likelihood~\cite{Planck-2018-cosmology}. Such neutrinos become nonrelativistic closer to recombination and change both the expansion rate around last scattering and the early ISW effect. The compression assumes the standard neutrino sector at these epochs, and neither the lensing spline nor the residual ISW template, which acts only on the largest scales, can mimic this change.
In the unlensed spectra this change resembles a slight tilt, which $n_\mathrm{s}$ absorbs.

We have also checked that, combined with DESI, the acoustic prior gives the same $w_0$--$w_a$ constraints as the full CMB likelihood from which it is derived.
Given the stability across late-time models and the agreement with the full likelihood, we provide the baseline posterior for direct use\footnote{A Cobaya implementation of the likelihood, with examples
for CAMB and CLASS, is available at
\url{https://github.com/JiangJQ2000/CMBAcousticPrior} and \url{https://JiangJQ2000.github.io/CMBAcousticPrior/}.}
as a multivariate Gaussian in $(100\theta_*,\omega_{\rm b},\omega_{\rm bc})$ with mean
\begin{equation}
 \boldsymbol{\mu}(100\theta_*,\omega_{\rm b},\omega_{\rm bc})=
 \begin{pmatrix}
  1.041003 \\
  0.02243963 \\
  0.14097667
 \end{pmatrix},
\end{equation}
and covariance matrix
\begin{equation}
 \boldsymbol{C}=10^{-9}\times
 \begin{pmatrix}
   44.434  & 4.1749  & -79.866 \\
   4.1749  & 9.8960  & -42.219 \\
  -79.866  & -42.219 & 1221.7
 \end{pmatrix}\,.
\end{equation}
The acoustic-scale convention used for $\theta_*$ and its CLASS counterpart are described in Appendix~\ref{app:compressed_covariance}.

The posterior on the compressed parameters serves as a prior for testing late-time physics, not as a CMB likelihood.
In practice the prior is evaluated at the $(100\theta_*,\omega_{\rm b},\omega_{\rm bc})$
predicted by each sampled late-time model and multiplied into the likelihood of the low-redshift data being analyzed. In models where the dark-matter density does not scale as $a^{-3}$ after recombination, $\omega_{\rm bc}$ in the prior is replaced by $\omega_{\rm bc}^*$ of \Cref{eq:omega_bc_star}, obtained by computing the baryon and cold dark matter density of the model at its own last-scattering epoch $a_*$ and scaling it to the present as $a^{-3}$. Because equality precedes these departures, $a_{\rm eq}$ needs no such rescaling, and the repository also provides the prior in $(100\theta_*,\omega_{\rm b},a_{\rm eq})$. Both steps assume the standard pre-recombination physics used to build the prior.
Since the compression does not retain the likelihood of the individual TT, TE, and EE measurements, its contribution to $\Delta\chi^2$ gives only a qualitative reading of how models compare with the CMB and does not quantify a $p$-value or significance. A frequentist comparison must instead use $\Delta\chi^2$ from the data likelihoods entering it, with the prior applied identically to both models. For nested models, such as $\Lambda$CDM within $w_0w_a$CDM, this $\Delta\chi^2$ can be converted into a $p$-value using the number of added parameters, provided the usual likelihood-ratio assumptions hold.
The acoustic prior can, however, be used directly in Bayesian analyses, both for parameter inference and for model comparison, where it enters the evidence of each model in the same way.

\section{Consistency of CMB lensing with late-time cosmological models}\label{sec:lensing}

\begin{figure*}[htp!]
 \centering
 \includegraphics[width=\textwidth]{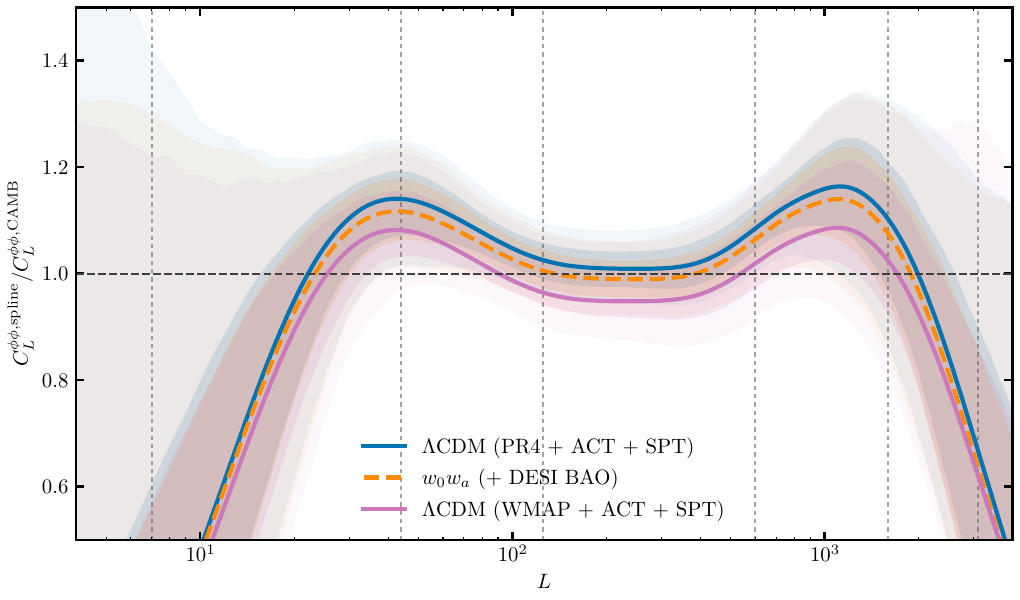}
 \caption{Reconstructed lensing-potential spectrum relative to the prediction of $\Lambda$CDM for PR4+ACT DR6+SPT-3G D1 and for WMAP+ACT DR6+SPT-3G D1, and of $w_0w_a$CDM with DESI (dashed). Curves give the best fit and bands the 68\% and 95\% credible intervals, with the spline nodes marked by vertical lines. The reconstruction exceeds the $\Lambda$CDM prediction near $L\simeq50$ and $L\sim10^3$, and the DESI-preferred $w_0w_a$ model closes part of this gap.}
 \label{fig:clpp_late_model_consistency}
\end{figure*}

\begin{figure*}[htp!]
 \centering
 \includegraphics[width=0.9\textwidth]{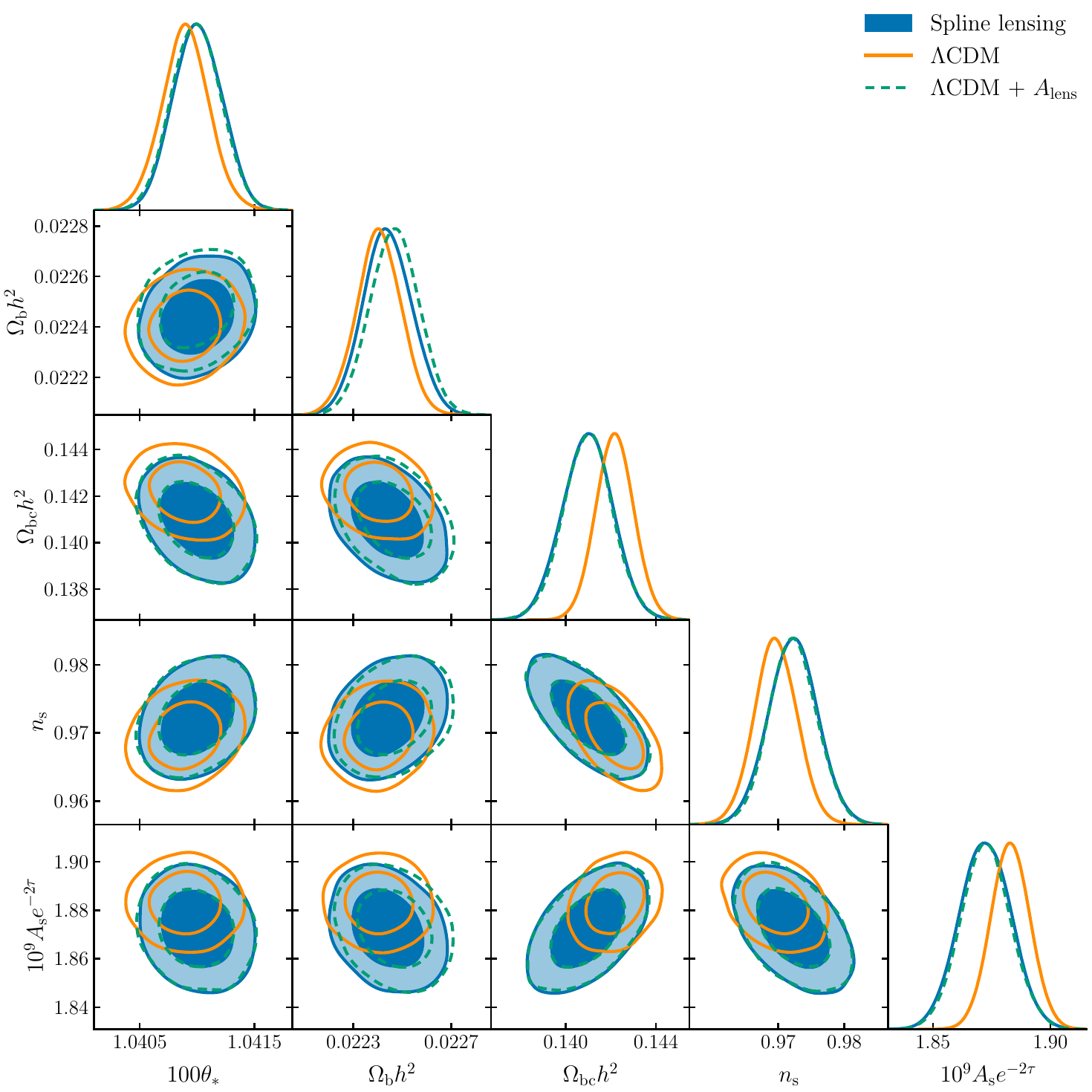}
 \caption{68\% and 95\% credible regions of the early-$\Lambda$CDM parameters from the baseline PR4+ACT DR6+SPT-3G D1 data under three treatments of the lensing potential. The filled contours use the spline reconstruction of $C_L^{\phi\phi}$ adopted for the compression, while the solid and dashed contours use the $\Lambda$CDM prediction without and with a free amplitude $A_{\rm lens}$, both with the full ISW contribution. Tying the lensing to $\Lambda$CDM raises $\omega_{\rm bc}$ and $A_\mathrm{s}e^{-2\tau}$, and freeing $A_{\rm lens}$ recovers the spline result apart from a small offset in $\omega_{\rm b}$.}
 \label{fig:lcdm_alens_spline}
\end{figure*}

The analysis above reconstructs the lensing-potential power spectrum, $C_L^{\phi\phi}$, in a late-Universe-agnostic way by replacing the model prediction with a flexible spline in amplitude. For each posterior sample, one can nevertheless impose a specific late-time cosmological model and compute the corresponding lensing spectrum from the matter power spectrum predicted by that model. Comparing these two spectra provides a direct diagnostic of whether the lensing smoothing in the primary TT, TE, and EE spectra and the lensing power measured by the four-point reconstruction require a departure from the lensing signal expected in a given late-time scenario.

Figure~\ref{fig:clpp_late_model_consistency} shows this comparison for two representative late-time models. In the first case, the late-time sector is fixed to $\Lambda$CDM with $\sum m_\nu=0.06$\,eV. The reconstructed spectrum is systematically higher than the $\Lambda$CDM prediction over the broad range of multipoles where the CMB data have the greatest constraining power. This behavior is consistent with the well-known lensing-excess preference in Planck temperature and polarization spectra, usually parameterized by a scale-independent phenomenological amplitude $A_{\rm lens}>1$~\cite{Calabrese:2008rt,Planck-2018-cosmology}.
The spline reconstruction further shows that, for the data combination considered here, the apparent excess may not be simply a uniform rescaling of the lensing spectrum.
Instead, it is driven by two localized enhancements, one around $L\simeq 50$ (between the second and third nodes) and one toward $L\sim 10^3$ (between the fourth and fifth nodes). Both exceed $2\sigma$ over a limited range of multipoles, and the higher-$L$ feature reaches a peak significance of $2.6\sigma$.
\footnote{This significance is quoted only as the pointwise deviation and does not account for the look-elsewhere effect or for correlations among neighboring multipoles.}
With WMAP in place of the Planck primary spectra, both features remain but at lower significance, $1.5\sigma$ near $L\simeq50$ and $1.3\sigma$ near $L\sim10^3$, while the reconstruction lies below the $\Lambda$CDM prediction between them. The two balance in the overall amplitude, $A_{\rm lens}=1.007\pm0.038$ compared with $1.047\pm0.030$ for the baseline, in line with the absence of excess lensing in the ACT DR6 spectra~\citep{AtacamaCosmologyTelescope:2025blo}.

Relative to $\Lambda$CDM, the spline and the residual ISW amplitude improve the best fit by $\Delta\chi^2\simeq-12.4$. About half of the improvement comes from the lensing reconstruction and most of the rest from the Planck and ACT spectra, so the four-point measurement and the peak smoothing pull in the same direction, although the free ISW amplitude may absorb part of the latter. A single scale-independent amplitude $A_{\rm lens}$ accounts for $3.0$ of this improvement, supplied by the peak smoothing in the ACT and Planck spectra, while the remaining $9.3$ comes mostly from the reconstruction. The higher-$L$ feature also lies above the baseline ACT range, $L\leq763$, where SPT-3G alone supplies lensing bandpowers. Because the locations of the features were identified after examining the reconstruction, and the gain is shared among seven added parameters that the data do not all constrain, their significance is difficult to quantify. We therefore regard them as hints that flag these scales for further study rather than as evidence for a departure from $\Lambda$CDM, and upcoming ground-based lensing measurements should determine whether they persist.

In Fig.~\ref{fig:lcdm_alens_spline} we show how the treatment of lensing feeds back on the early-$\Lambda$CDM parameters. When $C_L^{\phi\phi}$ follows the $\Lambda$CDM prediction, its amplitude is set by the matter density and the primordial amplitude, so the excess lensing preferred by the data is supplied mainly by raising $\omega_{\rm bc}$ and $A_\mathrm{s}e^{-2\tau}$. Both move up by about $1\sigma$ (and by less than $0.1\sigma$ for WMAP+ACT DR6+SPT-3G D1), with $n_\mathrm{s}$ moving down along the degeneracy that links the three in the primary spectra and $\theta_*$ and $\omega_{\rm b}$ moving down by up to $0.5\sigma$, and the errors on $\omega_{\rm bc}$ and $A_\mathrm{s}e^{-2\tau}$ shrink by about 25\% because the lensing now constrains them as well. This extra information holds only if the late-time growth is that of $\Lambda$CDM.
Freeing a single amplitude $A_{\rm lens}$ breaks the link and removes these shifts, with $A_{\rm lens}=1.05\pm0.03$.
Beyond this amplitude, the spline shifts $\omega_{\rm b}$, by $0.3\sigma$, and spends the rest of its freedom on the scale dependence of the lensing bandpowers that supplies most of the $\Delta\chi^2$ gain.

\begin{figure}[t]
 \centering
 \includegraphics[width=\columnwidth]{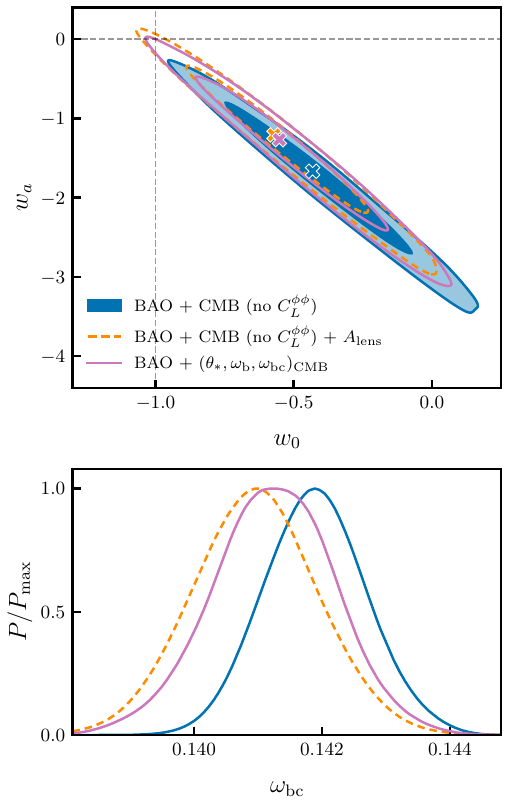}
 \caption{Effect of CMB lensing information on the $w_0w_a$ constraints from DESI.  The top panel compares the $w_0$--$w_a$ contours of the two-point CMB fit, the same fit with a free phenomenological smoothing amplitude $A_{\rm lens}$, and the acoustic prior.  Although the two-point fit uses no lensing reconstruction, its TT, TE, and EE spectra still carry lensing information through the peak smoothing predicted by the sampled cosmology.  Contours show the 68\% and 95\% credible regions, crosses the best-fit points, and dashed lines $w_0=-1$ and $w_a=0$.  The lower panel shows the corresponding marginalized posteriors of $\omega_{\rm bc}=\Omega_{\rm bc}h^2$.}
 \label{fig:desi_lensing_posterior}
\end{figure}

\section{The acoustic prior and dynamical dark energy}\label{sec:desi}

Motivated by recent DESI results, which show a preference for dynamical dark energy~\cite{DESI2024VI,DESI2024VII,DESI:2025zgx}, we consider the implications of the acoustic prior in this context using the Chevallier--Polarski--Linder parameterization for dark energy~\cite{Chevallier:2000qy,Linder:2002et},
\begin{equation}
 w(a)=w_0+w_a(1-a),
\end{equation}
where $w_0$ and $w_a$ are sampled together with the remaining cosmological parameters. Since CMB data alone only weakly constrain the $w_0w_a$ sector, we include the DESI measurements~\cite{DESI:2025zgx} to break the geometric degeneracy in the CMB distance to last scattering.
The orange band of Fig.~\ref{fig:clpp_late_model_consistency} compares the freely reconstructed lensing spectrum with the prediction of this $w_0w_a$ model.
Its prediction lies above that of $\Lambda$CDM and reduces part of the amplitude mismatch.

We compare the acoustic prior with a conventional fit to the two-point spectra that follows the ``CMB (no lensing)'' combination of Ref.~\cite{DESI:2025zgx}.  This fit uses the same high-$\ell$ spectra and SRoll2 low-$\ell$ EE as the baseline, adds the Planck low-$\ell$ TT likelihood~\cite{Planck-2018-likelihoods}, and omits the four-point lensing reconstruction.  Its TT, TE, and EE predictions are nonetheless lensed by the matter distribution of the sampled cosmology.  The two-point fit therefore retains late-time information through lensing-induced peak smoothing, whereas the acoustic prior marginalizes over that smoothing.  Comparing the two treatments isolates the impact of interpreting the smoothing within the sampled late-time model.

The two-point fit gives $w_0=-0.394^{+0.209}_{-0.244}$ and $w_a=-1.797^{+0.706}_{-0.576}$, placing $\Lambda$CDM outside the 95\% contour in \Cref{fig:desi_lensing_posterior} (at the 99.7\% credible level).  Replacing the two-point spectra with the acoustic prior shifts the result to $w_0=-0.491^{+0.208}_{-0.242}$ and $w_a=-1.465^{+0.721}_{-0.561}$, for which $\Lambda$CDM lies on the 95\% contour (at the 94.9\% credible level).  The best-fit point moves from $(w_0,w_a)=(-0.432,-1.666)$ in the two-point fit to $(-0.552,-1.267)$ in the compressed fit.
\footnote{We find that the current version of Ref.~\cite{DESI:2025zgx} inadvertently reported the value of $\Omega_mh^2$, which includes the contribution from massive neutrinos, as $\Omega_{bc}h^2$.  This makes their ``DESI + CMB (no lensing)'' result, based on the two-point CMB spectra, appear nearly identical to that obtained with the CMB acoustic prior.  After correcting this mismatch, even the PR4-only analysis exhibits a similar difference.}

The phenomenological $A_{\rm lens}$ fit provides a controlled check of the origin of this shift because, as shown above, the changes induced by $w_0w_a$CDM are broad rather than localized in scale.
Once the smoothing amplitude is decoupled from the cosmological prediction, the posterior becomes $w_0=-0.541^{+0.199}_{-0.238}$ and $w_a=-1.308^{+0.703}_{-0.538}$, close to the compressed result.  The residual smoothing preference is $A_{\rm lens}=1.062\pm0.036$, $1.7\sigma$ above unity.  Thus the additional displacement of the two-point contour is associated with its requirement that the DESI-selected late-time cosmology also reproduce the observed peak smoothing~\cite{Park2024,Peng:2025nez}.

The shift is also visible in the physical matter density.  The two-point fit yields $\omega_{\rm bc}=0.14189\pm0.00081$, compared with $0.14124\pm0.00095$ from the prior on $(\theta_*,\omega_{\rm b},\omega_{\rm bc})$, and the MAP values shift downward by a comparable amount.  Allowing $A_{\rm lens}$ to vary gives $\omega_{\rm bc}=0.14095\pm0.00099$, again close to the compressed result.  A larger $\omega_{\rm bc}$ raises the matter density available to produce CMB lensing, while the DESI-preferred $w_0w_a$ region can itself enhance the lensing potential through its altered expansion and growth history.  When the lensing spectrum is tied to the late-time model, these responses jointly pull the two-point fit along the $w_0$--$w_a$ degeneracy. Marginalizing the smoothing information or absorbing it into $A_{\rm lens}$ removes that extra pull.

\Cref{tab:w0wa_significance} quantifies the preference for $w_0w_a$CDM over $\Lambda$CDM for three treatments of the CMB, each combined with DESI alone and with Union3.1. When the lensing of the primary spectra follows the late-time model, as in the two-point fit, the preference is $2.97\sigma$, or $3.24\sigma$ with Union3.1. Freeing $A_{\rm lens}$ decouples the smoothing amplitude from that model and lowers these to $1.71\sigma$ and $2.37\sigma$. %
The spline fit used to build the acoustic prior, which also frees the shape of the lensing spectrum, gives a similar $1.88\sigma$ and $2.43\sigma$.
\begin{table}[t]
\centering
\caption{Preference for $w_0w_a$CDM over $\Lambda$CDM for three treatments of the CMB, combined with DESI and with DESI plus Union3.1, given as $\Delta\chi^2_{\rm MAP}$ and, in parentheses, the significance for two extra parameters. The spline fit, which also drops the late ISW effect and the low-$\ell$ temperature, is the likelihood behind the acoustic prior.
}
\label{tab:w0wa_significance}
\setlength{\tabcolsep}{7.5pt}\setlength{\aboverulesep}{0pt}\setlength{\belowrulesep}{0pt}
\begin{tabular}{lc|c}
\toprule
CMB variant & + DESI & + DESI + Union3.1 \\
\midrule
Model prediction & $11.7$ ($2.97\sigma$) & $13.4$ ($3.24\sigma$) \\
+ $A_{\rm lens}$ & $4.9$ ($1.71\sigma$) & $8.1$ ($2.37\sigma$) \\
Spline & $5.6$ ($1.88\sigma$) & $8.4$ ($2.43\sigma$) \\
\bottomrule
\end{tabular}
\end{table}

The same separation can be seen in distances through the matter-era distance interval (MEDI) introduced in Ref.~\cite{Weiner:2026MEDI},
\begin{equation}
 \frac{D_M(a_d,a_m)}{r_d}
 \equiv \frac{D_M(a_d)}{r_d}-\frac{D_M(a_m)}{r_d},
 \qquad a_m=\frac{1}{1+2.330}.
\end{equation}
Here $a_d=1/(1+z_{\rm drag})$ is the scale factor at the baryon drag epoch.
Within $\Lambda$CDM, the nongeometric information in the CMB spectra predicts this interval through the physical matter densities.  An independent acoustic-scale determination instead combines the CMB geometric constraint on $D_M(a_d)/r_d$ with $D_M(a_m)/r_d=38.85\pm0.12$, inferred in Ref.~\cite{Weiner:2026MEDI} from all DESI tracers in $\Lambda$CDM.

\begin{figure}[t]
 \centering
 \includegraphics[width=\columnwidth]{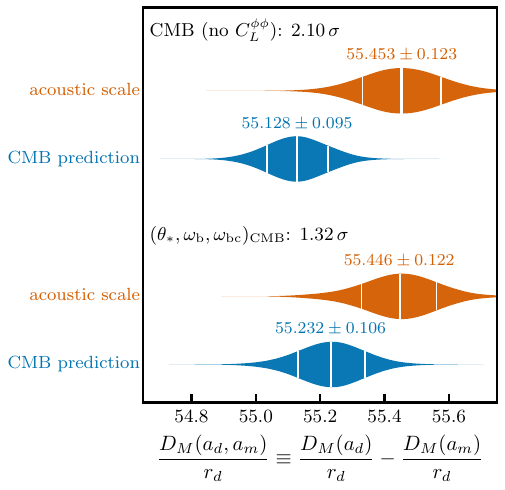}
 \caption{Matter-era distance interval predicted by the nongeometric CMB information (``CMB prediction'') and measured from the CMB acoustic scale together with all DESI tracers in $\Lambda$CDM (``acoustic scale'').  The upper pair uses the two-point CMB spectra without lensing reconstruction, while the lower pair uses the acoustic prior, in which the lensing-induced smoothing has been marginalized.  White lines mark the median and central 68\% interval, and the quoted values are posterior means and standard deviations.  The tensions are the differences in means divided by the quadrature sum of the standard deviations.}
 \label{fig:medi_lensing_comparison}
\end{figure}
\begin{figure*}[htp!]
\centering
\includegraphics[width=\textwidth]{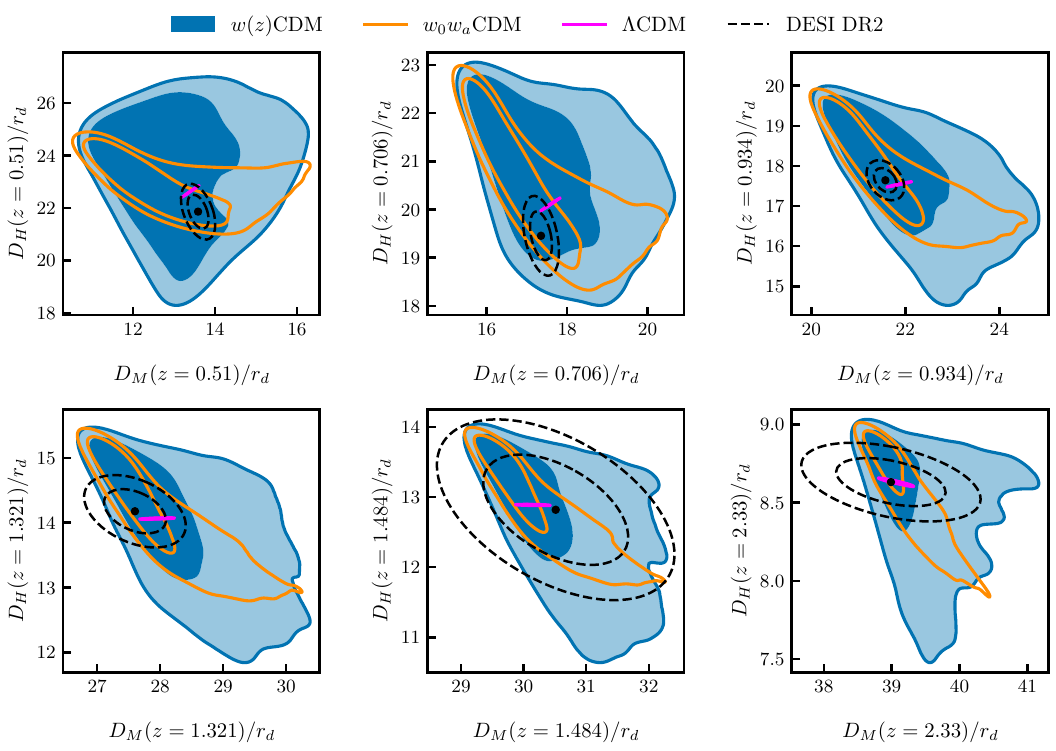}
\caption{The 68\% and 95\% posterior predictive regions of $D_M(z)/r_d$ and $D_H(z)/r_d$ at the six DESI tracer redshifts, given only the CMB constraints on $(\theta_*,\omega_{\rm b},\omega_{\rm bc})$, in the binned $w(z)$CDM (filled), $w_0w_a$CDM restricted to $w_0+w_a\ge-3$ like the binned prior (open), and $\Lambda$CDM (magenta), with the DESI measurements (dashed). Only $\Lambda$CDM is offset from DESI, at $z=0.51$ and $0.706$ and mainly in $D_H$, while the evolving models narrow in $D_M$ toward Ly$\alpha$, where dark energy is subdominant.}
\label{fig:cmb_desi_joint_distances}
\end{figure*}

For the two-point spectra, the two determinations differ by $2.10\sigma$ in \Cref{fig:medi_lensing_comparison}, and the acoustic prior reduces this to $1.32\sigma$.  The MEDI tension obtained from the two-point spectra therefore reflects the joint effect of the residual acoustic-scale mismatch and the lensing excess encoded by peak smoothing.  Since compression both shifts and broadens the CMB prediction, the change from $2.10\sigma$ to $1.32\sigma$ is not itself a separately defined significance for the lensing contribution.

The CMB prediction assumes $\Lambda$CDM, with a fixed neutrino mass, between the drag epoch and $z=2.33$.  The acoustic-scale determination assumes it at lower redshift, where it links the lower-redshift tracers to the Ly$\alpha$ measurement and makes the result 4.4 times more precise than the direct DESI value, $38.99\pm0.53$~\cite{DESI:2025zgx}.  Reducing DESI to a single transverse distance also hides which tracers, and which of $D_M$ and $D_H$, drive the comparison.

Given a class of late-time models, the CMB alone predicts every distance that DESI measures~\cite{Mortonson:2008qy,Mortonson:2009hk}. The MEDI comparison tests one such prediction, a single distance interval in $\Lambda$CDM. In \cref{fig:cmb_desi_joint_distances}, we extend this idea to every DESI tracer and to wider classes of late-time models, through the posterior predictive distributions of $D_M(z)/r_d$ and $D_H(z)/r_d=c/[H(z)r_d]$ in $\Lambda$CDM, $w_0w_a$CDM, and the binned $w(z)$CDM of \cref{sec:results}. Restricting to $\Lambda$CDM breaks the geometric degeneracy of \cref{sec:physics} by assumption, so it predicts each distance with sub-percent precision, while the $w_0w_a$CDM and binned $w(z)$CDM predictions stretch into elongated regions that reflect degeneracies among late-time expansion histories consistent with the acoustic constraints. These regions are broadest at low redshift because, as $z\to0$, $D_M(z)\to cz/H_0$ and $D_H(z)\to c/H_0$ depend only on $H_0$, which the CMB constrains weakly once dark energy evolves. This breadth comes from the wide prior range of $H_0$, not from a lack of information in the acoustic prior. Imposing a reasonable prior on $H_0$, such as $60\le H_0\le80\,{\rm km\,s^{-1}\,Mpc^{-1}}$, shrinks the low-redshift predictions substantially.

{\color{red}

}

{\color{red}

}

The $\Lambda$CDM prediction agrees with DESI from $z=0.934$ to Ly$\alpha$ but is offset at $z=0.51$ and $0.706$, where DESI prefers a smaller $D_H$. Physically, with $r_d$ and the matter density calibrated by the CMB, a smaller $D_H$ means a higher expansion rate at these redshifts than the cosmological constant fixed by $\theta_*$ allows. Once dark energy evolves, we find the DESI measurements within the predicted regions at every tracer redshift, for both $w_0w_a$CDM and the binned $w(z)$CDM.

The width of each prediction, which shrinks with redshift as dark energy becomes subdominant, sets what DESI can learn from it. The $\Lambda$CDM prediction is narrow enough that a single tracer can falsify it, as the measurements at $z=0.51$ and $0.706$ begin to do. Once dark energy is allowed to evolve, the $w_0w_a$CDM prediction is broad at these redshifts, so the same measurements instead constrain $w_0$ and $w_a$, as we found in \cref{fig:desi_lensing_posterior}. A future measurement outside the $w_0w_a$CDM prediction but inside that of the binned $w(z)$CDM, which can reshape $H(z)$ bin by bin, would trace the failure to the $w_0w_a$ form and leave the pre-recombination physics untouched. A measurement outside both would be hard to attribute to the late-time expansion, above all at Ly$\alpha$, where all three models predict $D_M/r_d$ to about $1\%$, and would point instead to the pre-recombination physics assumed in the CMB prediction, most directly through the calibration of $r_d$.

{\color{red}
}

\section{Discussion and Conclusions}\label{sec:conclusions}

We have presented an updated compression of the CMB using Planck PR4, ACT DR6, and SPT-3G D1 temperature and polarization measurements together with the joint CMB lensing reconstruction. The aim is to retain the information supplied by pre-recombination acoustic physics while minimizing assumptions about the subsequent expansion and growth history. We remove the dominant late-ISW sensitivity, marginalize over the residual contribution, and replace the cosmology-predicted lensing-potential spectrum with a flexible reconstruction constrained directly by the lensing data. The resulting CMB information is summarized by the three parameters $\{\theta_*,\omega_{\rm b},\omega_{\rm bc}\}$, determined to $0.02\%$, $0.4\%$, and $0.8\%$, respectively, which provide a compact calibration of the acoustic ruler for late-time analyses.

Relative to the previous Planck PR4 compression, the combined data shift $\omega_{\rm b}$ and $n_\mathrm{s}$ upward by $1.4\sigma$ and $1.6\sigma$ in units of the PR4 errors, while $\theta_*$ and $\omega_{\rm bc}$ move by less than $0.5\sigma$. The individual CMB data sets (Planck PR4, ACT DR6 and SPT-3G D1) are also mutually consistent after marginalizing over late-time physics, supporting their combination into one acoustic prior. Freeing the neutrino mass shifts $\omega_{\rm b}$ by $0.2\sigma$, while the dark energy extensions shift the three parameters by well below $0.1\sigma$. Even with $w$ free in five redshift bins, the errors on the three parameters do not grow. With no measurable late-time information left in the compression, one prior serves all of these late-time models. For models in which the dark-matter density does not scale as $a^{-3}$ after recombination, the prior should be evaluated at the recombination-era density $\omega_{\rm bc}^*$ of \Cref{eq:omega_bc_star} rather than at its present-day value.

While this work was in preparation, two related compressions of the CMB appeared~\cite{Srivastav:2026toe,Pedrotti:2026ncd}. Unlike our construction, neither removes the late-time lensing and ISW contributions before compressing, so the compressed parameters retain the response of the cosmology in which they were calibrated, and \citet{Srivastav:2026toe} accordingly supplies a separate compression for each model class. \citet{Pedrotti:2026ncd} concludes that, beyond dark-energy extensions, the full CMB likelihood should be the default. Their compression is calibrated in $\Lambda$CDM on primary spectra that remain lensed and include the low-$\ell$ temperature, so dropping the lensing reconstruction leaves this late-time information in place, and \cref{sec:lensing} shows that the lensing alone pulls $\omega_{\rm bc}$ upward. The full likelihood they use as the benchmark predicts the lensing and ISW from each tested model, so it is not agnostic to the physics under test. A mismatch with it therefore need not mean the compression fails, since the relevant test for a late-time prior is invariance under the late-time model, which ours passes in all five models of \cref{fig:late}, interacting dark energy included.

Treating the lensing potential independently of late-time physics also allows us to examine the origin of the lensing excess without reducing it to a single phenomenological amplitude. The reconstruction exceeds the $\Lambda$CDM prediction in two localized features, near $L\simeq50$ and $L\sim10^3$, the latter peaking at $2.6\sigma$. Both weaken when WMAP replaces the Planck primary spectra. When the lensing is instead tied to the $\Lambda$CDM prediction, the excess is supplied by raising $\omega_{\rm bc}$ and $A_\mathrm{s}e^{-2\tau}$, information that holds only if the late-time growth is that of $\Lambda$CDM. The reconstruction absorbs the excess in $C_L^{\phi\phi}$ itself, so it no longer pulls $\omega_{\rm bc}$ upward.

This distinction matters most for evolving dark energy, because the treatment of lensing changes the significance with which DESI and the CMB prefer it. When the lensing of the primary spectra follows the late-time model, the preference for $w_0w_a$CDM over $\Lambda$CDM is $2.97\sigma$, rising to $3.24\sigma$ with Union3.1 (\cref{tab:w0wa_significance}). A free $A_{\rm lens}$ lowers these to $1.71\sigma$ and $2.37\sigma$. The likelihood behind our acoustic prior, which also frees the shape of the lensing spectrum and drops the late ISW effect and the low-$\ell$ temperature, gives $1.88\sigma$ and $2.43\sigma$. The preference therefore has a geometric part that survives without tying the lensing to a late-time model and a lensing part that the standard analysis adds. The lensing part is also the more model dependent, because the lensing predicted for a given $w(a)$ depends on how the dark energy affects the growth of structure, a choice to which the acoustic prior remains agnostic.

We have also extended the idea behind the matter-era distance interval, a single comparison in $\Lambda$CDM, to every DESI distance and to any late-time model, since the acoustic prior predicts each distance without BAO input once that model is chosen (\cref{fig:cmb_desi_joint_distances}). A measurement outside the $\Lambda$CDM prediction but inside the $w_0w_a$CDM one points to evolving dark energy, one outside $w_0w_a$CDM but inside the binned $w(z)$CDM to a richer expansion history, and one outside all three to the pre-recombination physics that the prior assumes is standard. DESI points toward the first class, offset from $\Lambda$CDM at $z=0.51$ and $0.706$, mainly in $D_H$. At Ly$\alpha$, where dark energy is subdominant, all three models fix $D_M/r_d$ to about $1\%$ and agree with DESI, a clean check of the acoustic physics itself. Because the predictions come from the CMB alone, any future BAO measurement can be read against them directly, and those at higher redshift test the pre-recombination physics most directly. Models that alter the dark matter before recombination lie outside this test.

Our results illustrate the utility of separating the primary acoustic scale from the late-time growth information encoded by CMB lensing. A self-consistent cosmological model should ultimately describe both, but when testing departures from $\Lambda$CDM it is useful to distinguish the information that calibrates the early Universe standard ruler from that which depends on the subsequent growth history. The acoustic prior provides such a separation and offers a complementary tool for testing late-time cosmological models with reduced dependence on assumptions about CMB lensing and growth.
Forthcoming temperature and polarization data from SPT-3G over a quarter of the sky~\cite{SPT-3G:2025mju} and from the Simons Observatory~\cite{SimonsObservatory:2018koc} will extend precise measurements deep into the damping tail, where much of the information in the acoustic prior resides. In the SPT-3G forecast, much of the gain in $\omega_{\rm c}$ comes from interpreting the lensing reconstruction within $\Lambda$CDM~\cite{SPT-3G:2025mju}. In the acoustic prior that reconstruction constrains the shape of $C_L^{\phi\phi}$ instead, so the gain in $\omega_{\rm bc}$ will depend on how tightly it confines the spline, while the lensing itself decides whether the features near $L\simeq50$ and $L\sim10^3$ survive.

\acknowledgments
We thank Antony Lewis and Dhiraj Kumar Hazra for helpful comments and feedback on the manuscript. The authors acknowledge the use of the high-performance computing clusters at the Korea Astronomy and Space Science
Institute. K.L. acknowledges support from the U.S. Department of Energy under Grant No. DE-SC0016021. J.J., P.M., and A.S. acknowledge funding from the Korea Astronomy and Space Science Institute (KASI) through the project ``Research on the Principles of the Accelerating Expansion of the Universe'' (Project Code: 2026183201). The authors acknowledge the use of Claude (Anthropic) to improve the clarity and readability of the manuscript and take full responsibility for its content.

\bibliographystyle{mod-apsrev4-2}
\bibliography{references,DESI_supporting_papers}

\appendix

\section{The acoustic-scale convention in CAMB and CLASS}
\label{app:compressed_covariance}
\label{app:camb_class}

CAMB and CLASS define the acoustic angle $\theta_*=r_s(z_*)/D_M(z_*)$
differently.
CAMB evaluates it at the redshift where the Thomson optical depth, excluding
reionization, reaches unity~\cite{Lewis:1999bs}.
CLASS~\cite{class_2011arXiv1104.2932L} instead provides
\texttt{theta\_s\_100}, evaluated at the maximum of the visibility function,
and \texttt{theta\_star\_100}, evaluated where the total Thomson optical depth,
including reionization, reaches unity.
The latter matches the CAMB definition only when reionization is switched off
with \texttt{reio\_parametrization = reio\_none}.

\begin{figure}[htp!]
 \centering
 \includegraphics[width=\columnwidth]{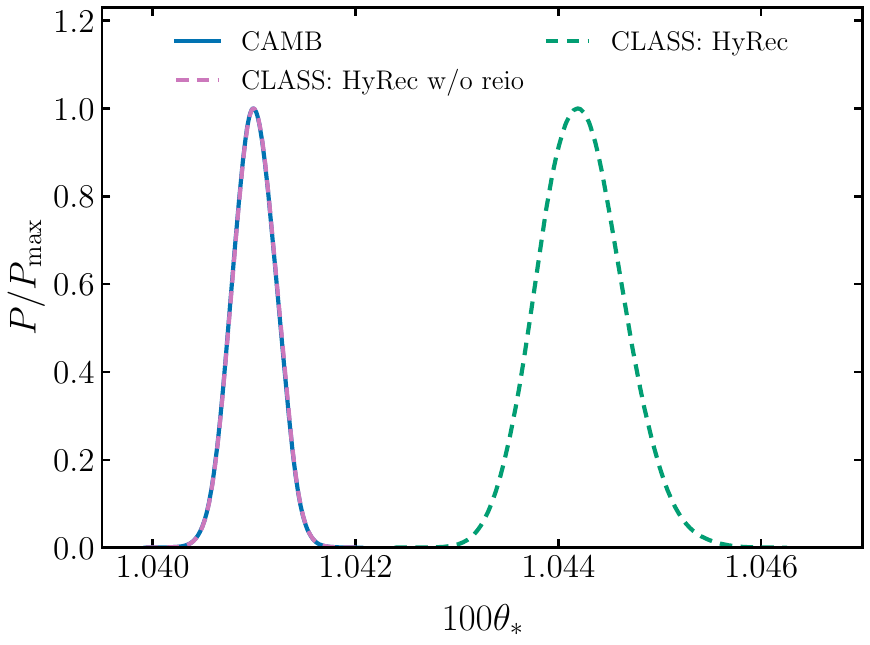}
 \caption{Marginalized posterior of $100\theta_*$ from CAMB (blue), compared
 with the value recalculated with CLASS for each sample of the CAMB chain.
 With reionization switched off and the settings listed in the text
 (pink dashed), CLASS reproduces the CAMB
 posterior. Under the default reionization treatment (green dashed) the acoustic scale is
 displaced by about $15\sigma$.
 }
 \label{fig:camb_class_thetastar}
\end{figure}

The mean and covariance given in \Cref{sec:results} are computed with CAMB,
taking $100\theta_*$ from the derived parameter \texttt{thetastar} with
\texttt{recombination\_model = CosmoRec}.
To compare the two codes we recalculate \texttt{theta\_star\_100} with CLASS
for every sample in the CAMB chain, retaining the original sample weights.
In Fig.~\ref{fig:camb_class_thetastar} we compare the CAMB posterior of
$100\theta_*$ with the values recalculated by CLASS, with reionization
switched off and at its default setting.
With reionization switched off, one massive neutrino of $0.06$\,eV with
\texttt{N\_ur = 2.0308} so that $N_{\rm eff}=3.044$, \texttt{T\_cmb = 2.7255},
and the helium abundance from the PRIMAT-based BBN table
\texttt{sBBN\_2025\_primat.dat}, we find that CLASS reproduces the CAMB
posterior, even though it uses
HyRec~\cite{Ali-Haimoud:2010hou} in place of CosmoRec~\cite{Chluba:2010fy}.
The agreement is sensitive to the radiation content, since $N_{\rm eff}=3.046$
already shifts $100\theta_*$ by $0.34\sigma$, whereas leaving reionization at
its default setting displaces it by about $15\sigma$.

A single prior therefore serves both codes.
CLASS users need only switch off reionization when computing
\texttt{theta\_star\_100} and adopt the radiation and neutrino content listed
above, which the prior assumes, whereas the choice of BBN helium table has a
negligible effect on $100\theta_*$.
The mean and covariance of \Cref{sec:results} then apply without modification.

\section{The role of the low-\texorpdfstring{$\ell$}{l} EE likelihood}
\label{app:lowell_ee}

Our baseline analysis includes the Planck SRoll2 low-$\ell$ EE likelihood to constrain the optical depth to reionization.  To test whether this choice affects the early Universe compression, we repeat the full Planck PR4+ACT DR6+SPT-3G D1 analysis after removing SRoll2, while leaving the primary high-$\ell$ spectra, the joint lensing-reconstruction likelihood, cosmological model, spline freedom, and priors unchanged.

The resulting constraints are compared in Table~\ref{tab:lowell_ee}.  Removing low-$\ell$ EE has a negligible effect on the acoustic prior. The means of $\theta_*$, $\omega_{\rm b}$ and $\omega_{\rm bc}$ shift by less than $0.06\sigma$, and their uncertainties change by at most $2\%$. Among the remaining parameters, $n_\mathrm{s}$ shifts the most, by less than $0.2\sigma$.

\begin{table}[htp!]
  \centering
  \caption{Marginalized 68\% constraints from the baseline dataset and without the Planck SRoll2 low-$\ell$ EE likelihood.}
  \label{tab:lowell_ee}
  \begin{tabular}{@{}lc@{\hspace{18pt}}c@{}}
    \toprule
    Parameter & Baseline & Without low-$\ell$ EE \\
    \midrule
    $100\theta_*$ & $1.04100\pm0.00021$ & $1.04100\pm0.00022$ \\
    $\omega_{\rm b}$ & $0.022440\pm0.000099$ & $0.022440\pm0.000099$ \\
    $\omega_{\rm bc}$ & $0.1410\pm0.0011$ & $0.1409\pm0.0011$ \\
    $n_\mathrm{s}$ & $0.9722\pm0.0037$ & $0.9729\pm0.0038$ \\
    $10^9 A_\mathrm{s}e^{-2\tau}$ & $1.872\pm0.011$ & $1.871\pm0.011$ \\
    \bottomrule
  \end{tabular}
\end{table}

As expected, low-$\ell$ EE remains essential for separating the primordial amplitude from reionization. Removing SRoll2 replaces the baseline constraint $\tau=0.0570^{+0.0053}_{-0.0065}$ with the one-sided limit $\tau<0.169$ at 95\% confidence and broadens the marginalized constraint on $A_\mathrm{s}$ correspondingly.  The high-$\ell$ spectra, however, constrain the combination $A_\mathrm{s}e^{-2\tau}$ directly.  Its stability, together with that of the acoustic scale, baryon density, matter density, and spectral tilt, demonstrates that the five-parameter early Universe compression does not rely sensitively on the low-$\ell$ EE likelihood or on its assumed reionization history.

\section{Results with alternative priors on the lensing potential spectrum}
\label{app:lensing_spline_priors}

Our baseline reconstruction uses six nodes at
$L=\{7,44,125,600,1600,3100\}$ and independent flat priors
$p_i\equiv\log\!\left([L_i(L_i+1)]^2C_{L_i}^{\phi\phi}/2\pi\right)
\in[-22,-14]$.
We test the sensitivity of our results to both the resolution and the
parameterization of this reconstruction.
First, we increase the number of nodes to 11 by adding nodes at
$L=\{18,74,274,980,2227\}$ and simultaneously widen every node prior to
$p_i\in[-24,-12]$.
Second, retaining the six baseline node positions, we instead spline the
logarithmic ratio to the lensing spectrum predicted by CAMB at each sampled
cosmology,
\begin{equation}
 q_i=\log\!\left(C_{L_i}^{\phi\phi}/
 C_{L_i}^{\phi\phi,\mathrm{CAMB}}\right),
 \qquad q_i\in[-3,3].
 \label{eq:alternative_clpp_prior}
\end{equation}
In the latter parameterization, $q_i=0$ is the exact model-predicted lensing
spectrum, whereas the broad prior permits order-of-magnitude departures from
it.

\begin{figure*}[htp!]
 \centering
 \includegraphics[width=\textwidth]{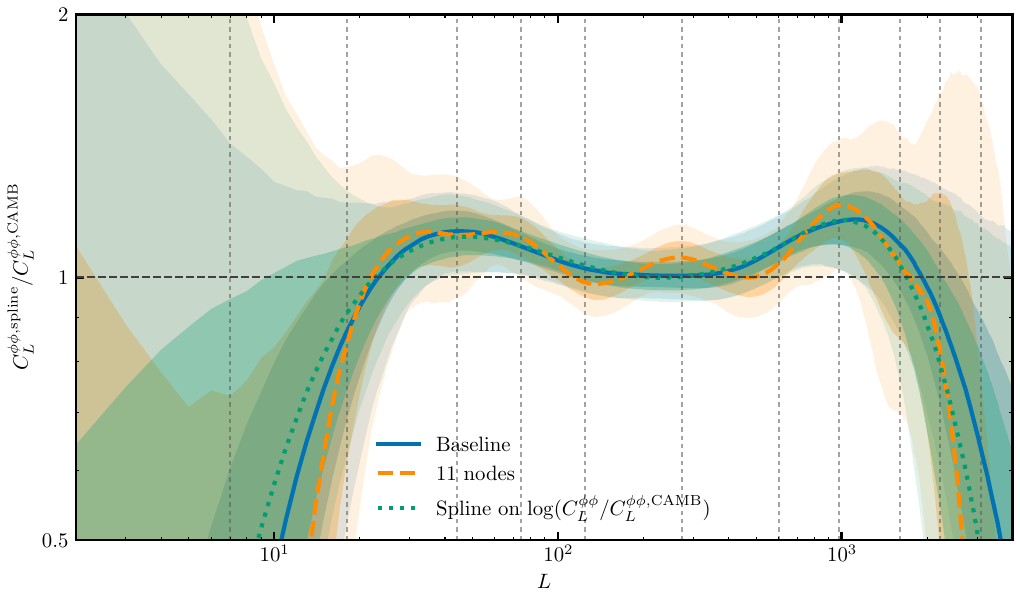}
 \caption{Robustness of the reconstructed lensing-potential power spectrum to
 the spline resolution and prior parameterization.  We compare the six-node
 baseline reconstruction, an 11-node reconstruction with wider priors on the
 absolute node amplitudes, and a six-node spline in
 $\log(C_L^{\phi\phi}/C_L^{\phi\phi,\mathrm{CAMB}})$.  The solid curves show
 posterior medians, while the dark and light bands denote the 68\% and 95\%
 credible intervals, respectively.  Vertical dotted lines mark the union of
 the node positions used by the reconstructions.}
 \label{fig:clpp_prior_robustness}
\end{figure*}

As shown in Fig.~\ref{fig:clpp_prior_robustness}, the three reconstructions
agree over the multipole range constrained by the data.  In particular, the
enhanced lensing power around $L\sim50$--$100$ and the rise near
$L\sim10^3$ both persist when the spline resolution is increased and when the
prior is recentered on the CAMB prediction.  The posterior broadens toward
the lowest and highest multipoles, where the data have less constraining
power and the spline extrapolation is consequently more prior-dependent.

The inferred early Universe parameters are even less sensitive to these
choices.  Relative to the baseline chain, the posterior means of
$\{\theta_*,\omega_{\rm b},\omega_{\rm bc},n_\mathrm{s},
A_\mathrm{s}e^{-2\tau}\}$ shift by at most $0.09$ times the corresponding
baseline posterior standard deviation in the 11-node analysis, and their
standard deviations change by at most $1.5\%$.  For the spline-ratio
parameterization, the corresponding maximum changes are $0.038$ standard
deviations and $3.4\%$.  The five-parameter early Universe constraints are
therefore insensitive to the lensing-spline resolution and to these broad
alternative priors.

\end{document}